\documentclass[a4paper,11pt]{article}

\usepackage{jheppub} 
\usepackage{lineno}
\usepackage{dcolumn} 
\usepackage{epsfig}
\usepackage{graphicx}
\usepackage{subcaption}

\usepackage{mathtools}
\usepackage{amsmath}
\usepackage{amssymb, bm}
\usepackage{mathrsfs}
\usepackage[capitalise]{cleveref} 
\usepackage{slashed}
\usepackage{physics}

\usepackage[usenames, dvipsnames]{xcolor}
\usepackage{hyperref}
\usepackage{multirow}
\usepackage{array}
\usepackage[T1]{fontenc}
\usepackage[normalem]{ulem}

\title{Exploring nuclear modification using one-point energy correlator
at the electron-ion collider}

\author[a]{Yu Fu,}
\author[b,c,d]{Zhong-Bo Kang,}
\author[b,c]{Jani Penttala,}
\author[e,f,d]{Yiyu Zhou}

\affiliation[a]{Department of Physics, Duke University, Durham, NC 27708, USA}
\affiliation[b]{Department of Physics and Astronomy, University of California, Los Angeles, CA 90095, USA}
\affiliation[c]{Mani L. Bhaumik Institute for Theoretical Physics, University of California, Los Angeles, CA 90095, USA}
\affiliation[d]{Center for Frontiers in Nuclear Science, Stony Brook University, Stony Brook, NY 11794, USA}
\affiliation[e]{Department of Physics, University of Turin, via Pietro Giuria 1, I-10125 Torino, Italy}
\affiliation[f]{INFN, Section of Turin, via Pietro Giuria 1, I-10125 Torino, Italy}

\emailAdd{yu.fu@duke.edu}
\emailAdd{zkang@physics.ucla.edu}
\emailAdd{janipenttala@physics.ucla.edu}
\emailAdd{yiyu.zhou@unito.it}

\abstract{
We study the one-point energy correlator (OPEC) at both the back-to-back and collinear limits in electron-proton and electron-nucleus collisions.
We provide the factorization formalism for the two types of OPEC and present phenomenological predictions in the kinematic region relevant for the future Electron-Ion Collider.
Focusing on cold nuclear matter effects in electron-nucleus scattering, we demonstrate that the OPEC serves as a powerful probe of the transverse momentum dependent physics and in characterizing the medium-induced transverse momentum broadening in cold nuclear matter.
}
\arxivnumber{2512.16847}

\begin{document}
\maketitle
\flushbottom

\section{Introduction}

In high-energy collisions, event-shape observables \cite{Dasgupta:2003iq} quantify the geometric and dynamical properties of final-state energy flows, providing a global characterization of the event topology.
They have long played a central role in advancing our understanding of Quantum Chromodynamics (QCD).
In recent years, the advent of high-precision measurements and jet substructure \cite{Larkoski:2017jix, Kogler:2018hem, Marzani:2019hun} techniques has enabled a more local and differential exploration of energy flow within individual jets, giving rise to observables that resolve the internal structure of the collimated sprays of particles.

A particularly powerful and unifying framework to study both global and local aspects of energy flow is provided by the energy correlators \cite{Moult:2025nhu}, which measure the energy correlation as a function of the relative angle between the detectors.
From the field-theoretical perspective, energy-flow correlators are defined in terms of light-ray energy flow operators along direction $\vec{n}$:
\begin{align}
\mathcal{E} \pqty{\vec{n}}
=
\int_0^{\infty} \dd{t}
\lim_{r\to\infty}
r^2 n^i T_{0i} \pqty{t, r \vec{n}}
,
\end{align}
where the $T_{\mu\nu}$ is stress-energy tensor \cite{Sveshnikov:1995vi, Hofman:2008ar}.
The two-point energy correlator (or energy-energy correlator, EEC) belongs to a classic type of observables in the energy correlator family, which are defined as:
\begin{align}
\mathrm{EEC} \pqty{\chi}
=
\int \dd{\vec{n}_1} \dd{\vec{n}_2}
\frac{\expval{\mathcal{E} \pqty{\vec{n}_1} \mathcal{E} \pqty{\vec{n}_2}}}{Q^2}
\delta \pqty{\vec{n}_1 \cdot \vec{n}_2 - \cos{\chi}}
,
\end{align}
where $Q = \sum_{i \in X} E_i$ is the total energy in the phase-space region $X$ over which the correlator is defined, and $\chi$ is the opening angle between the two detectors.
The EEC was first introduced in the 1970s as a clean test of perturbative QCD \cite{Basham:1978bw, Basham:1978zq}.
Modern collider experiments, with exceptional angular resolution, have enabled precise studies of energy correlators across various collision systems: from elementary collisions \cite{Komiske:2022enw, Holguin:2023bjf, CMS:2024mlf, ALICE:2024dfl, STAR:2025jut} to those involving either cold \cite{Devereaux:2023vjz, Fu:2024pic, Barata:2024wsu, Nambrath:2025ttz} or hot nuclear medium \cite{Andres:2022ovj, Andres:2023xwr, Barata:2023bhh, Yang:2023dwc, Xing:2024yrb, Ke:2025ibt}.
Depending on the measurement setup, the EEC can serve either as a global event-shape observable or as a jet-substructure probe, of which the latter is sensitive to the angular distribution of energy within reconstructed jets.
On the theoretical side, substantial progress has been achieved in understanding the QCD factorization and resummation properties of the EEC in both the back-to-back and collinear limits \cite{Moult:2018jzp, Dixon:2019uzg}.

While two-point correlators probe pairwise energy correlations, a simpler yet equally informative observable can be constructed by correlating the energy flow with respect to a fixed direction, and that is the one-point energy correlator (OPEC).
Formally, it is defined as:
\begin{align}
\mathrm{OPEC}_{n_0} \pqty{\chi}
=
\int \dd{\vec{n}}
\frac{\expval{\mathcal{E} \pqty{\vec{n}}}}{Q}
\delta \pqty{\vec{n} \cdot \vec{n}_0 - \cos{\chi}}
,
\end{align}
where $\vec{n}_0$ denotes a chosen reference direction, such as the beam axis or the jet axis in high-energy particle collisions.
In the context of scattering processes, the expectation value of the energy flow operator can be expressed as:
\begin{align}
\expval{\mathcal{E} \pqty{\vec{n}}}
=
\frac{1}{\sigma}\sum_i \frac{\dd{\sigma_i}}{\dd{\vec{n}}} E_i\,,
\end{align}
normalized by the cross section $\sigma$, where $\sigma_i$ represents the inclusive cross section for producing hadron $i$ with energy $E_i$. 
The summation over all final-state hadrons weighted by their energies renders the OPEC infrared and collinear safe.
Originally proposed in the 1970s \cite{Basham:1977iq}, the OPEC has recently attracted renewed interest as a versatile tool for probing transverse momentum dependent (TMD) dynamics, the internal structure of the nucleon and parton hadronization~\cite{Li:2021txc, Kang:2023big, Mi:2025abd, Gao:2025evv, Song:2025bdj, Gao:2025cwy,Zhu:2025qkx, Cao:2025icu}, among which the correlation of energy flows at the back-to-back limit ($\chi \to \pi$) and the collinear limit ($\chi \to 0$) are of particular importance.
As we will see, the simplicity and angular sensitivity of the OPEC make it particularly suitable for studying intrinsic transverse momentum distributions.

Despite recent developments in understanding OPEC, much less is known about the behavior of these quantities under the presence of a nuclear medium.
Understanding nuclear modification is essential for unraveling the dynamics of in-medium scattering and transverse momentum broadening in cold nuclear matter.
The upcoming Electron-Ion Collider (EIC)  \cite{Accardi:2012qut, AbdulKhalek:2021gbh} offers a unique opportunity to explore these effects in a clean and controllable environment, where one can extract cold nuclear effects by comparing electron-proton ($e+p$) and electron-nucleus ($e+A$) collisions.
The study of OPEC in cold nuclear environment offers a new, differential probe of transverse momentum broadening, complementing the picture of traditional TMD measurements which does not have energy weighting \cite{Bacchetta:2006tn, Kang:2017glf}.

In this work, we investigate the observable OPEC in both the back-to-back and collinear limits, with a particular focus on extending its study from proton to nuclear targets.
We first consider the deep inelastic scattering (DIS) processes in the kinematic region where the final hadron and incoming beam are almost back-to-back in the Breit frame.
The OPEC considered here is defined as the correlation between the energy flow carried by the final hadron and the energy flow carried by the incoming beam.
The study of hadronic energy in the final state in SIDIS was explored in~\cite{Meng:1991da,Nadolsky:1999kb}.
In recent years, this observable was studied in~\cite{Li:2021txc},  where its back-to-back limit was investigated within the TMD framework, and the full angular range was explored using the event generator \textsc{Pythia} \cite{Bierlich:2022pfr}.
This analysis was later extended to polarized DIS to probe TMD structures through the azimuthal dependence of OPEC \cite{Kang:2023big}.
However, these studies were restricted to $e+p$ collisions.
In this work, we extend the investigation to $e+A$ systems to explore cold-nuclear-matter effects.

In addition, we analyze lepton-jet production in the back-to-back configuration \cite{Liu:2018trl, Liu:2020dct}, focusing on the OPEC in the collinear limit, where the energy flow is measured inside the produced jet.
Here, by collinear limit, we refer to the OPEC that measures the correlation between the energy carried by a hadron within the jet and the total jet energy that flows along the jet axis. The observable measures the energy-weighted density of particles as a function of their distance from the jet axis.
Therefore, it is equivalent to the differential jet shape \cite{Seymour:1997kj, Li:2011hy, Chien:2014nsa, Kang:2017mda, Neill:2018wtk, Cal:2019hjc, Ke:2024emw}.
As noticed in \cite{Kang:2017mda}, jet shapes admit a description in terms of TMD fragmentation functions.
In this work, our goal is to establish a TMD factorized formula for calculating OPEC observables at the back-to-back and collinear limits.
Within this formalism, the OPEC factorizes into a hard function and non-perturbative components encoded in TMD parton distribution functions (TMD PDFs) and EEC jet functions, among which the latter is constructed from TMD fragmentation functions (TMD FFs).
Cold nuclear matter effects can be systematically incorporated by modeling the corresponding nuclear TMD PDFs (nTMD PDFs) and nuclear TMD FFs (nTMD FFs).
Based on this factorization formalism, we investigate nuclear modification systematically and provide phenomenological predictions for $e+p$ and $e+A$ collisions.

The structure of this paper is as follows.
In \cref{sec:b2b}, we investigate the OPEC between back-to-back final-state hadrons and the incoming proton or nucleus in DIS.
After introducing the relevant kinematics in \cref{sec:b2b-kinematics}, we present the factorization formula for this observable within the TMD framework in \cref{sec:b2b-factorization}, and discuss in detail its essential components: the quark TMD PDFs and the EEC jet function.
We then carry out a phenomenological study to illustrate the nuclear modification of the OPEC by comparing numerical results in $e+p$ and $e+A$ collisions.
In \cref{sec:collinear}, we perform a similar analysis for the OPEC in the collinear limit, where we examine the hadronic energy flow inside the jet produced in $e+p$ and $e+A$ collisions.
Based on the corresponding TMD factorized formula, we study the nuclear effects by comparing the results of $e+p$ and $e+A$ collisions, and provide phenomenological implications for future measurements at the EIC.
We finally summarize our findings in \cref{sec:conclusion}.

\section{One-point energy correlator at back-to-back limit}
\label{sec:b2b}

In this section, we study the OPEC in the DIS processes in the kinematic region where the final hadron and initial proton beam are almost back-to-back in the Breit frame.
We first set up the kinematics and then provide a factorized formula for the OPEC within the TMD framework.
This is followed by a detailed discussion on the important ingredients in the formalism: TMD PDFs and EEC jet functions.
Finally, we provide the numerical results for the phenomenological study of this observable at the future EIC.

\subsection{Kinematics}
\label{sec:b2b-kinematics}

\begin{figure}
\centering
\includegraphics[width = 0.7 \linewidth]{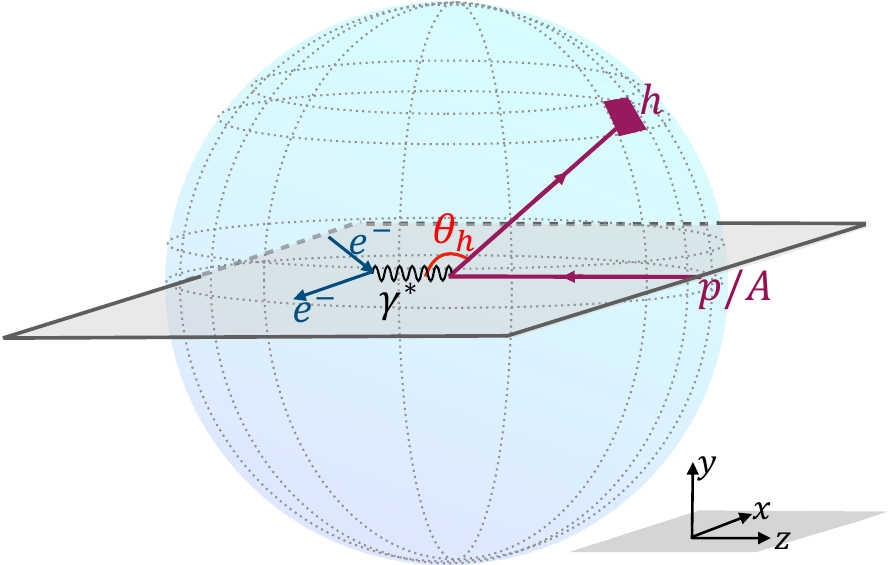}
\caption{
Illustration of OPEC for DIS in the Breit frame, where the incoming proton or nucleus ($p$/$A$) moves along the $-z$ axis and the virtual photon $\gamma^*$ propagates along the $+z$ direction.
Together with the incoming lepton, they define the lepton plane (shown in gray).
The angle between the detector that measures hadron $h$ and the incoming proton or nucleus is denoted as $\theta_h$.
}
\label{fig:SIDIS}
\end{figure}

For the process $e \pqty{\ell} + p/A \pqty{P} \to e \pqty{\ell'} + h \pqty{P_h} + X$, as illustrated in \cref{fig:SIDIS}, we define the usual semi-inclusive deep inelastic scattering (SIDIS) kinematic variables:
\begin{align}
x
\equiv
\frac{Q^2}{2P \cdot q}
, \quad
y
\equiv
\frac{Q^2}{x s}
, \quad
z_h
\equiv
\frac{P \cdot P_h}{P \cdot q}
, 
\end{align}
which are the Bjorken $x$, the event inelasticity and the final hadron's momentum fraction, respectively.
Here, $q \equiv \ell - \ell'$ is the momentum of the exchanged virtual photon with virtuality $Q^2 \equiv -q^2$ and $s \equiv \pqty{\ell + P}^2$ is the electron-nucleon center-of-mass energy.
We also denote the corresponding opening angle of a given final-state hadron measured with respect to the proton beam direction as $\theta_h$.
For convenience, we also often use the variable $\tau$ defined as:
\begin{align}
\tau
\equiv
\frac{1 + \cos{\theta_h}}{2}
=
\cos[2](\frac{\theta_h}{2})
.
\end{align}

In the limit where the incoming proton/nucleus and the detector are almost back-to-back, \textit{i.e.}, $\theta_h \to \pi$ or $\tau \to 0$, up to power corrections, we have:
\begin{align}
\tau
=
\frac{\boldsymbol{P}_{hT}^2}{z_h^2 Q^2}
=
\frac{\boldsymbol{q}_{T}^2}{Q^2}
, \label{eq:tau_and_qT}
\end{align}
for a given outgoing hadron inside the detector. Here, $\bm{P}_{hT}$ is the transverse momentum of the final hadron $h$.
For convenience, we have also introduced the shorthand notation
$\bm{q}_T \equiv -\bm{P}_{hT}/z_h$. In the Breit frame used here,
$\bm{q}_T$ should be considered as an auxiliary variable.%
\footnote{The quantity $\boldsymbol{q}_T$ is equal to the transverse momentum of the virtual photon in the so-called hadron-hadron frame, where the incoming proton and final-state hadron define the longitudinal axis.
More discussion of the relation between the two frames can be found in section ~2.6 of Ref.~\cite{Boussarie:2023izj}.}

\subsection{Factorization formalism}
\label{sec:b2b-factorization}

The OPEC in SIDIS can be defined as:
\begin{align}
\frac{\dd{\Sigma_{\mathrm{SIDIS}}}}{\dd{x} \dd{Q^2} \dd{\tau}}
=
\sum_h
\int_0^\pi \dd{\theta_h} \int_0^1 \dd{z_h}
z_h \frac{\dd{\sigma_{\mathrm{SIDIS}}}}{\dd{x} \dd{Q^2} \dd{z_h} \dd{\theta_{h}} }
\delta \pqty{\tau -
\frac{1+\cos{\theta_h}}{2}}
.
\end{align}
Under the back-to-back limit of the incoming proton/nucleus and the observed hadron, we can apply the relation between the final hadron's transverse momentum and $\tau$ as given in \cref{eq:tau_and_qT}, and relate the OPEC to the $\bm{q}_{T}$-differential cross section in SIDIS:
\begin{align}
\frac{\dd{\Sigma^{\mathrm{b.t.b.}}}}{\dd{x} \dd{Q^2} \dd{\tau}}
=
\sum_h
\int \dd[2]{\bm{q}_T}
\int_0^1 \dd{z_h}
z_h \frac{\dd{\sigma_{\mathrm{SIDIS}}}}{\dd{x} \dd{Q^2} \dd{z_h} \dd[2]{\bm{q}_T}}
\delta \pqty{\tau -
\frac{\bm{q}_T^2}{Q^2}}
. \label{eq:EECDIS_qT}
\end{align}
The factorization of $\bm{q}_T$-differential cross section for the unpolarized SIDIS is given by \cite{Gourdin:1972kro, Kotzinian:1994dv, Diehl:2005pc, Bacchetta:2006tn, Anselmino:2008sga, Collins:2011zzd, Kang:2015msa, Kang:2023big}:
\begin{align}
\frac{\dd{\sigma_{\mathrm{SIDIS}}}}{\dd{x} \dd{Q^2} \dd{z_h} \dd[2]{\bm{q}_T}}
& =
\frac{\sigma_0}{xs}
H (Q,\mu)
\int_0^\infty \frac{b \dd{b}}{2 \pi}
\sum_q
e_q^2 J_0 \pqty{b q_T}
f^{(u)}_q \pqty{x, b, \mu, \frac{\zeta}{\nu^2}}
\nonumber \\
& \qquad \qquad \qquad \qquad \qquad \times
z_h^2
D^{(u)}_{h/q} \pqty{z_h,b,\mu,\frac{{\zeta}'}{\nu^2}}
S_{n n_h} \pqty{b, \mu, \nu}
,
\end{align}
where $f^{(u)}_q \pqty{x, b, \mu, \zeta/\nu^2}$ and $D^{(u)}_{h/q} \pqty{z_h,b,\mu,{\zeta}'/\nu^2}$ are the Fourier transformed ``unsubtracted'' quark TMD PDFs and TMD FFs defined as:
\begin{align}
f_q^{(u)} \pqty{x, b, \mu, \frac{\zeta}{\nu^2}}
& =
\int \dd[2]{\bm{p}_T}
e^{-i \bm{p}_T \cdot \bm{b}}
f_q^{(u)} \pqty{x, p_T, \mu, \frac{\zeta}{\nu^2}}
, \\
D_{h/q}^{(u)} \pqty{z_h, b, \mu, \frac{{\zeta}'}{\nu^2}}
& =
\frac{1}{z_h^2}
\int \dd[2]{\bm{k}_T}
e^{-i \bm{k}_T \cdot \bm{b} / z_h}
D^{(u)}_{h/q} \pqty{z_h, k_T, \mu, \frac{{\zeta}'}{\nu^2}}
, \label{eq:FF-FT}
\end{align}
where $\bm{p}_T$ is the parton's transverse momentum with respect to its parent hadron
and $\bm{k}_T$ is the transverse momentum of the fragmented hadron relative to its parent parton.
Conventionally, we also define $p_T \equiv \abs{\bm{p}_T}$ and $k_T \equiv \abs{\bm{k}_T}$.

The Born cross section is given by:
\begin{align}
\sigma_0
=
\frac{2 \pi \alpha_{\mathrm{em}}^2}{Q^2}
\frac{1+(1-y)^2}{y}
,
\label{eq:hard-func}
\end{align}
where $\alpha_{\mathrm{em}}$ is the fine structure constant.
At next-to-leading order (NLO), we also have the hard function that is given by \cite{Liu:2018trl, Arratia:2020nxw, Ellis:2010rwa}:
\begin{align}
H \pqty{Q, \mu}
=
1
+
\frac{\alpha_s}{2\pi} C_F
\bqty{-\ln[2](\frac{\mu^2}{Q^2}) - 3 \ln(\frac{\mu^2}{Q^2}) - 8 + \frac{\pi^2}{6}}
,
\label{eq:hard-func-NLO}
\end{align}
with $\alpha_s$ being the strong coupling constant, and we will choose $\mu=Q$ for our phenomenology studies. The contribution to the hard function at even higher orders can be found in \textit{e.g.} Refs.~\cite{Gehrmann:2005pd, Gehrmann:2010ue, Abele:2021nyo, Bonino:2024qbh, Bonino:2024wgg, Goyal:2024emo, Bonino:2025qta, Bonino:2025bqa, Goyal:2023zdi,Goyal:2024tmo,Goyal:2025qyu}.

The soft function $S_{n n_h}$ encodes the contribution from soft gluon radiation and can be computed using a rapidity regulator~\cite{Chiu:2011qc}. Its result differs from the standard soft function $S_{n \overline{n}}$ through a rescaling of the rapidity scale $\nu$, yielding~\cite{Li:2020bub, Fang:2023thw}:
\begin{align}
S_{n n_h} \pqty{b, \mu, \nu}
=
S_{n \overline{n}} \pqty{b, \mu, \nu \sqrt{\frac{n \cdot n_h}{2}}}
, \label{e.soft-function-n-n_h}
\end{align}
where $n \cdot n_h = 1 - \tanh{\eta_h}$ with $\eta_h$ being the rapidity of the final-state hadron, and $S_{n \overline{n}}$ denotes the standard soft function that is given by \cite{Collins:2011zzd, Kang:2021ffh}:
\begin{align}
S_{n  \overline{n}}(b,\mu,\nu)
=
1
& -
\frac{\alpha_s C_F}{2\pi}
\Bigg[
2 \pqty{\frac{2}{\eta} + \ln(\frac{\nu^2}{\mu^2})}
\pqty{\frac{1}{\epsilon} + \ln(\frac{\mu^2}{\mu_b^2})}
\nonumber \\
& \qquad \qquad \quad
+
\ln[2](\frac{\mu^2}{\mu_b^2})
-
\frac{2}{\epsilon^2}
+
\frac{\pi^2}{6}
\Bigg]
, \label{eq:Snnbar}
\end{align}
where we adopt dimensional regularization in $4-2\epsilon$ space-time dimensions and the rapidity regulator $\eta$ \cite{Chiu:2012ir}, and $\mu_b \equiv 2e^{-\gamma_E}/b$ with $\gamma_E$ being the Euler-Mascheroni constant.

By further defining the ``unsubtracted'' EEC jet function for quark \cite{Moult:2018jzp} as:
\begin{align}
J_q^{(u)} \pqty{b, \mu, \frac{{\zeta}'}{\nu^2}}
\equiv
\sum_h
\int_0^1 \dd{z_h}
z_h^3 D^{(u)}_{h/q} \pqty{z_h, b, \mu, \frac{{\zeta}'}{\nu^2}}
. \label{eq:unpjet}
\end{align}
we can write the OPEC at the back-to-back limit in SIDIS as:
\begin{align}
\frac{\dd{\Sigma^{\mathrm{b.t.b.}}}}{\dd{x} \dd{Q^2} \dd{\tau}}
=
\frac{Q^2}{2xs}
\sigma_0
H (Q,\mu)
\int_0^{\infty} \dd{b} b
&
\sum_q e_q
J_0 \pqty{b \sqrt{\tau Q^2}}
f_q^{(u)} \pqty{x, b, \mu, \frac{\zeta}{\nu^2}}
\nonumber \\
& \times
J_q^{(u)} \pqty{b, \mu, \frac{{\zeta}'}{\nu^2}}
S_{n  n_h} \pqty{b, \mu, \nu}
. \label{eq:EECDIS_unsub}
\end{align}
We note that the factorization formula above is written in terms of the ``unsubtracted'' TMD quantities (the TMD quark distribution and the EEC jet function), and they contain rapidity divergences.
In the following sections, we will identify all the components appearing in the factorization formula and systematically remove the rapidity divergences by subtracting the soft contributions from the TMD distributions.
This procedure yields the ``subtracted'' TMD distributions $f_q$ and $J_q$ that are defined later in \cref{e.subtracted-TMD-PDFs-definition} and \cref{e.subtracted-TMD-FFs-definition}, respectively.
Using these subtracted TMD distributions, one will find that the factorization formula above can be rewritten in a fully physical form:
\begin{align}
\frac{\dd{\Sigma^{\mathrm{b.t.b.}}}}{\dd{x} \dd{Q^2} \dd{\tau}}
=
\frac{Q^2 \sigma_0}{2xs}
H (Q,\mu)
\int_0^{\infty} \dd{b} b
\sum_q e_q
J_0 \pqty{b \sqrt{\tau Q^2}}
f_q \pqty{x, b, \mu, \zeta}
J_q \pqty{b,\mu,\hat{\zeta}}
, \label{eq:EECDIS_sub}
\end{align}
where $\hat{\zeta} = {\zeta}' \pqty{n \cdot n_h/2}^2$ and it will be discussed in \cref{sec:EEC-jet-function}.
Following the renormalization group (RG) consistency condition, we choose $\zeta = \hat{\zeta} = Q^2$ (see \cref{ss.RG-consistency-at-b-to-b-limit} for more details).

\subsubsection{Quark Distribution Function}
\label{sec:Quark-Distribuition-function}

The ``unsubtracted'' TMD quark distribution function $f_q^{(u)} \pqty{x, b, \mu, \zeta/\nu^2}$ contains the Collins-Soper scale $\zeta$ \cite{Collins:2011zzd, Ebert:2019okf, Boussarie:2023izj} and a rapidity renormalization scale $\nu$ \cite{Chiu:2012ir}.
The rapidity divergence in $f_q^{(u)}$ can be canceled by absorbing a square root of the standard soft function $S_{n  \overline{n}} \pqty{b, \mu, \nu}$ whose expression at the NLO is given in \cref{eq:Snnbar}.
Hence, we can define the ``subtracted'' TMD PDFs \cite{Collins:2011zzd}:
\begin{align}
f_q \pqty{x, b, \mu, \zeta}
=
f_q^{(u)} \pqty{x, b, \mu, \frac{\zeta}{\nu^2}}
\sqrt{S_{n  \overline{n}} (b, \mu, \nu)}
, \label{e.subtracted-TMD-PDFs-definition}
\end{align}
and it is free of the rapidity divergence.

TMD evolution for the ``subtracted'' TMD PDFs is governed by two equations: the Collins-Soper-Sterman evolution associated with the Collins-Soper scale $\zeta$ \cite{Collins:2011zzd, Boussarie:2023izj} and the renormalization group equation related to the scale $\mu$.
They are given by:
\begin{align}
\frac{\dd{}}{\dd{\ln{\sqrt{\zeta}}}}
\ln{f_q \pqty{x, b, \mu, \zeta}}
& =
K \pqty{b, \mu}
, \label{eq:zeta_evl} \\ 
\frac{\dd{}}{\dd{\ln{\mu}}}
\ln{f_q \pqty{x, b, \mu, \zeta}}
& =
\gamma_{\mu}^q \bqty{\alpha_s \pqty{\mu}, \frac{\zeta}{\mu^2}}
, \label{eq:mu_evl}
\end{align}
where $K \pqty{b, \mu}$ is the Collins-Soper evolution kernel \cite{Collins:2011zzd, Moult:2022xzt, Duhr:2022yyp, Boussarie:2023izj} and the $\mu$-evolution kernel is:
\begin{align}
\gamma^q_{\mu} \bqty{\alpha_s(\mu), \frac{\zeta}{\mu^2}}
=
-C_F \Gamma_{\mathrm{cusp}} \bqty{\alpha_s(\mu)}
\ln(\frac{\zeta}{\mu^2})
+
\gamma_q \bqty{\alpha_s \pqty{\mu}}
.
\end{align}
Here, $\Gamma_{\mathrm{cusp}}$ and $\gamma_q$ are the cusp and non-cusp anomalous dimensions, respectively.
Their expansions are given by:
\begin{align}
\Gamma_{\mathrm{cusp}} \bqty{\alpha_s \pqty{\mu}}
& =
\sum_{n=1} \Gamma_{n-1} \pqty{\frac{\alpha_s}{4\pi}}^n
, \\
\gamma_q \bqty{\alpha_s \pqty{\mu}}
& =
\sum_{n=1} \gamma_{n-1}^q \pqty{\frac{\alpha_s}{4\pi}}^n
.
\end{align}
For the above expansion we keep the following terms \cite{Korchemsky:1987wg, Becher:2006mr, Jain:2011xz}:
\begin{align}
\Gamma_0
& =
4
, \quad
\gamma_0^q
=
6 C_F
, \quad
\Gamma_1
=
C_A \pqty{\frac{268}{9} - \frac{4 \pi^2}{3}}
-
\frac{80}{9} T_F n_f
,
\end{align}
where $C_F = 4/3$, $C_A = 3$, $T_F = 1/2$ and $n_f$ is the number of flavors.
Solving the renormalization group equations in both $\mu$ and $\zeta$, while consistently incorporating the non-perturbative effects that arise in the large-$b$ region ($b \gg 1/\Lambda_{\mathrm{QCD}}$), yields the expressions for TMD PDFs:
\begin{align}
f_q \pqty{x, b, \mu, \zeta}
=
f_q \pqty{x, b, \mu_{b_*}, \mu_{b_*}^2}
\exp[-S_{\mathrm{pert}} \pqty{\mu, \mu_{b_*}, \zeta}]
\exp[-S_{\mathrm{NP}}^f \pqty{b, Q_0, \zeta}]
,
\end{align}
where the perturbative evolution from initial scales ($\mu_0, \zeta_0$) to final scales ($\mu, \zeta$) is encoded in $S_{\mathrm{pert}}$, whereas the non-perturbative (NP) contributions from large $b$ are parameterized in the function $S_{\mathrm{NP}}^f$.
We have chosen the initial scales as $\mu_0 = \sqrt{\zeta_0} = \mu_{b_*}$, for which we adopted $\mu_{b_*}$-prescription \cite{Collins:1984kg, Sun:2014dqm, Echevarria:2020hpy, Isaacson:2023iui} and defined:
\begin{align}
\mu_{b_*}
\equiv
\frac{2 e^{-\gamma_E}}{b_*}
, \quad
b_*
\equiv
\frac{b}{\sqrt{1 + b^2/b^2_{\max}}}
,
\end{align}
with $b_{\max} = 1.5~\mathrm{GeV}^{-1}$.
The perturbative Sudakov factor $S_{\mathrm{pert}} \pqty{\mu, \mu_{b_*}, \zeta_i}$ is given by:
\begin{align}
S_{\mathrm{pert}} \pqty{\mu, \mu_{b_*}, \zeta}
=
- K \pqty{b_*, \mu_{b_*}}
\ln(\frac{\sqrt{\zeta}}{\mu_{b_*}})
-
\int_{\mu_{b_*}}^{\mu} \frac{\dd{\mu'}}{\mu'}
\gamma^q_{\mu} \bqty{\alpha_s \pqty{\mu'}, \frac{\zeta}{\mu^{\prime 2}}}
. \label{eq:Spert}
\end{align}
Throughout this paper, we work at the next-to-leading logarithmic (NLL) level in resummation accuracy, where $K \pqty{b_*, \mu_{b_*}} = 0$.
On the other hand, we follow \cite{Sun:2014dqm, Echevarria:2020hpy} and parameterize the NP kernel as:
\begin{align}
S_{\mathrm{NP}}^f \pqty{b, Q_0, \zeta}
=
\frac{g_2}{2} \ln(\frac{b}{b_*}) \ln(\frac{\sqrt{\zeta}}{Q_0})
+
g_1^q b^2
, \label{e.S_NP-f-parameterization}
\end{align}
with $Q_0^2 = 2.4~\mathrm{GeV}^2$, $g_2 = 0.84$ and $g_1^q = 0.106~\mathrm{GeV}^2$.

In the conventional TMD approach for small-$b$ \cite{Boussarie:2023izj}, one can express $f_q \pqty{x, b, \mu_{b^*}, \mu_{b^*}^2}$ in terms of the collinear PDFs $f_q \pqty{x, \mu}$ through operator product expansion (OPE):
\begin{align}
f_q \pqty{x, b, \mu_{b_*}, \mu_{b_*}^2}
=
\sum_j \int_x^1 \frac{\dd{x'}}{x'}
C_{q \leftarrow j} \pqty{\frac{x}{x'},\mu_{b_*}}
f_j \pqty{x', \mu_{b_*}}
\equiv
\bqty{C_{q \leftarrow j} \otimes f_j} \pqty{x, \mu_{b_*}}
,
\label{eq:PDF-OPE}
\end{align}
where the matching coefficients $C_{q \leftarrow j}$ are perturbatively calculable and can be found in \cite{Aybat:2011zv, Collins:2011zzd, Kang:2015msa, Echevarria:2016scs, Luo:2019szz, Echevarria:2020hpy, Luo:2020epw, Ebert:2020yqt}.
In this work, we choose the coefficient function at the leading order $C_{q \leftarrow j} \pqty{x,b} = \delta_{qj} \delta \pqty{1-x}$.
Therefore, the TMD PDFs now read:
\begin{align}
f_q \pqty{x, b, \mu, \zeta}
=
f \pqty{x, \mu_{b_*}}
\exp[-S_{\mathrm{pert}} \pqty{\mu, \mu_{b_*}, \zeta}]
\exp[-S_{\mathrm{NP}}^f \pqty{b, Q_0, \zeta}]
. \label{eq:fq_para}
\end{align}

To incorporate the nuclear effects for $e+A$ collisions, the TMD PDFs should be replaced by the nuclear TMD (nTMD) PDFs.
To this end, we identify two sources of the modification originating from the nuclear effects.
First, in the OPE of the TMD PDFs in \cref{eq:PDF-OPE}, the collinear PDFs in vacuum should be replaced by their nuclear counterparts.
In the phenomenological analysis throughout this work, we adopt the EPPS16 nuclear PDF set \cite{Eskola:2016oht} for Au.
For the proton PDFs, we use the CT14nlo set \cite{Dulat:2015mca} to maintain consistency with the baseline employed in EPPS16.
Second, we follow the work in \cite{Alrashed:2021csd} and introduce a nuclear modified non-perturbative parameter $g_{1,A}^q$ in place of $g_1^q$ in \cref{e.S_NP-f-parameterization}:
\begin{align}
g_{1,A}^q
=
g_1^q + a_N \pqty{A^{1/3} - 1}
,
\end{align}
where $A$ is the mass number of the nucleus and the fit parameter $a_N = 0.016~\mathrm{GeV}^2$ characterizes the nuclear broadening.

\subsubsection{EEC Jet Function}
\label{sec:EEC-jet-function}

The unpolarized EEC quark jet function has been defined in \cref{eq:unpjet}, which is a definition that relies on the ``unsubtracted'' TMD FFs.
In order to remove the rapidity divergence in the ``unsubtracted'' TMD FFs, we again define the ``subtracted'' TMD FFs:
\begin{align}
D_{h/q} \pqty{z, b, \mu, \hat{\zeta}}
=
D^{(u)}_{h/q} \pqty{z, b, \mu, \frac{\hat{\zeta}'}{\nu^2}}
\frac{S_{n  n_h} \pqty{b, \mu, \nu}}{\sqrt{S_{n \overline{n}} \pqty{b, \mu, \nu}}}
, \label{e.subtracted-TMD-FFs-definition}
\end{align}
where $\hat{\zeta} = \hat{\zeta}' \pqty{n \cdot n_h/2}^2$, $S_{n  n_h}$ is the soft function given in \cref{e.soft-function-n-n_h} and $S_{n \overline{n}}$ is the standard soft function given in \cref{eq:Snnbar}.
We choose $\zeta = \hat{\zeta} = Q^2$ for the TMD distributions involved in the factorization formalism~\cite{Kang:2023oqj,Fang:2024auf}.
Solving the QCD evolution for the ``subtracted'' TMD FFs, we find:
\begin{align}  
D_{h/q} \pqty{z, b, \mu, \hat{\zeta}}
& =
D_{h/q} \pqty{z, b, \mu_{b_*}, \mu_{b_*}^2}
\exp[-S_{\mathrm{pert}} \pqty{\mu, \mu_{b_*}, \hat{\zeta}}]
\nonumber \\
& \quad \times
\exp[-S_{\mathrm{NP}}^D \pqty{z, b, Q_0, \hat{\zeta}}]
, \label{eq:TMDFFtoFF}
\end{align}
where $S_{\mathrm{pert}} \pqty{\mu, \mu_{b_*}, \hat{\zeta}}$ is the perturbative Sudakov factor given in \cref{eq:Spert}, and it resums the global logarithms from both rapidity and $\mu$-evolution.
The parameterization of the non-perturbative Sudakov factor is given by \cite{Kang:2017glf}:
\begin{align}
S_{\mathrm{NP}}^D \pqty{z, b, Q_0, \hat{\zeta}}
=
\frac{g_2}{2} \ln(\frac{b}{b_*}) \ln(\frac{\sqrt{\hat{\zeta}}}{Q_0})
+
g_1^D \frac{b^2}{z^2}
, \label{e.S_NP-FFs}
\end{align}
where $g_2 = 0.84$, $Q_0^2 = 2.4~\mathrm{GeV}^2$ and $g_1^D = 0.042~\mathrm{GeV}^2$ following from \cite{Echevarria:2020hpy, Sun:2014dqm}.
The TMD FFs can be perturbatively matched onto the collinear FFs $D_{h/i} \pqty{z, \mu}$ via an OPE in the limit of small $b$ as:
\begin{align}
D_{h/q} \pqty{z, b, \mu, \hat{\zeta}}
& =
\frac{1}{z^2}
\sum_i \int_z^1 \frac{\dd{y}}{y}
\hat{C}_{i \leftarrow q} \pqty{\frac{z}{y}, b}
D_{h/i} \pqty{y,\mu_{b_*}}
\nonumber \\
& \quad \times
\exp[-S_{\mathrm{pert}} \pqty{\mu, \mu_{b_*}, \hat{\zeta}}]
\exp[-S_{\mathrm{NP}}^D \pqty{z, b, Q_0, \hat{\zeta}}]
, \label{eq:FF-OPE}
\end{align}
where the matching coefficients $\hat{C}_{i \leftarrow q}$ can be found in Refs. \cite{Luo:2019szz, Echevarria:2020hpy, Luo:2020epw, Ebert:2020yqt}.
In this work, we use the leading order matching coefficients $\hat{C}_{i \leftarrow q} \pqty{z, b} = \delta_{iq} \delta \pqty{1-z}$.
Therefore, the ``subtracted'' EEC quark jet function can be written as:
\begin{align}
J_q \pqty{b, \mu, \hat{\zeta}}
& =
\sum_h
\int_0^1 \dd{z}
z^3 D_{h/q} \pqty{z, b, \mu, \hat{\zeta}}
\nonumber \\
& =
\sum_h
\int_0^1 \dd{z}
z \, D_{h/q} \pqty{z, \mu_{b_*}}
\exp(-g_1^D \frac{b^2}{z^2})
\nonumber \\
& \quad \times
\exp[-S_{\mathrm{pert}} \pqty{\mu, \mu_{b_*}, \hat{\zeta}}]
\exp[-\frac{g_2}{2} \ln(\frac{b}{b_*}) \ln(\frac{\sqrt{\hat{\zeta}}}{Q_0})]
. \label{eq:EECjetfunction}
\end{align}
For numerical convenience and stability, in the phenomenological analysis that follows, we represent the above $z$ integration by a fitted function \cite{Li:2021txc, Kang:2023oqj, Kang:2024otf}:
\begin{align}
\sum_h
\int_0^1 \dd{z}
z\, D_{h/q} \pqty{z, \mu_{b_*}}
\exp(-g_1^D \frac{b^2}{z^2})
\equiv
\exp[-S_{\mathrm{NP}}^{\mathrm{EEC}} \pqty{b}]
,
\end{align}
with the functional form:
\begin{align}
S_{\mathrm{NP}}^{\mathrm{EEC}} \pqty{b}
=
N b^{\alpha} \pqty{1 + rb^{\beta}}
. \label{eq:SEEC_NP_para}
\end{align}
To enforce a strictly larger exponent in the second term, we impose the constraints $\alpha > 0$ and $\beta > 0$.
Here, the DSS 2021 parameterization \cite{Borsa:2021ran} was adopted for the collinear FFs $D_{h/q}(z,\mu)$, including neutral and charged pions.
The fitted parameters in a vacuum environment are given in \cref{tab:nuclear_params}, as in Ref.~\cite{Kang:2024otf}. Finally, the EEC jet function is given by:
\begin{align}
J_q \pqty{b, \mu, \hat{\zeta}}
=
\exp[-S_{\mathrm{pert}} \pqty{\mu, \mu_{b_*}, \hat{\zeta}}]
\exp[-\frac{g_2}{2} \ln(\frac{b}{b_*}) \ln(\frac{\sqrt{{\hat{\zeta}}}}{Q_0}) - S_{\mathrm{NP}}^{\mathrm{EEC}} \pqty{b}]
. \label{eq:EECJF_para}
\end{align}

\begin{table}[t]
\centering
\begin{tabular}{l|cccc}
\hline
& $\alpha$ & $\beta$ & $N \bqty{\mathrm{GeV}^{\alpha}}$ & $r \bqty{\mathrm{GeV}^{\beta}}$ \\
\hline
proton & 0.42 & 1.00 & 0.45  & 0.63 \\
gold & $0.56 \pm 0.01$ & $1.21 \pm 0.02$ & $1.21 \pm 0.02$ & $0.138 \pm 0.008$ \\
\hline
\end{tabular}
\caption{
Values of fitted parameters from \cref{eq:SEEC_NP_para}.
The average and standard deviation values for gold are calculated from 29 sets of parameters corresponding to the Hessian sets given in LIKEn 2021 FFs \cite{Zurita:2021kli}.
}
\label{tab:nuclear_params}
\end{table}

To account for nuclear effects in $e+A$ collisions, the TMD FFs should be replaced by their nuclear correspondence.
In this way, the corresponding nuclear modifications are naturally incorporated into the EEC jet function.
The effects of these nuclear modifications are manifested in two key respects.
First, in the OPE of the TMD FFs in \cref{eq:FF-OPE}, the collinear FFs in vacuum should be replaced by the nuclear collinear FFs.
In this work, the LIKEn 2021 set \cite{Zurita:2021kli} for gold (Au) is used.
Second, we follow the
work in \cite{Alrashed:2021csd} and introduce a modified non-perturbative parameter $g_{1,A}^D$ in place of $g_{1}^D$ in \cref{e.S_NP-FFs}:
\begin{align}
g_{1,A}^D
=
g_1^D + b_N \pqty{A^{1/3}-1}
,
\end{align}
where $A$ is the mass number of the nucleus and $b_N = 0.0097~\mathrm{GeV}^2$ is the fitted parameter provided in \cite{Alrashed:2021csd}.
The fitted parameters for the case of Au are listed in \cref{tab:nuclear_params}, with average and standard deviation values computed from 29 sets of parameters provided in the LIKEn 2021 set \cite{Zurita:2021kli}.
We remark that, in our framework, the relevant nuclear effects in the final-state process are effectively encoded in the nTMD FFs.
This treatment is motivated by Ref.~\cite{Alrashed:2021csd}, where the medium-modified TMD FFs were fitted to existing $e+A$ SIDIS and $p+A$ Drell-Yan data and were shown to describe the data well.
\subsection{Phenomenology}

In this section, we make numerical predictions for the OPEC at the back-to-back region in DIS, based on the factorization formalism \labelcref{eq:EECDIS_sub} established in the previous sections.
Our focus will be on the nuclear modification of the OPEC.
To this end, we compare the results from $e+\mathrm{Au}$ collision to those from $e+p$ collision.

In our numerical calculations, we adopt the highest center-of-mass energy currently envisioned in the EIC design for electron-nucleus collisions \cite{AbdulKhalek:2021gbh}, namely $\sqrt{s} = 90~\mathrm{GeV}$.
The virtuality of the exchanged photon is fixed at $Q^2 = 20~\mathrm{GeV}^2$.
At the EIC, the inelasticity variable $y$ typically spans the range $0.01 \leq y \leq 0.95$.
Under this kinematic setup, the corresponding Bjorken $x$ values lie approximately within $2.5 \times 10^{-3} \lesssim x \lesssim 2.5 \times 10^{-1}$.
In this work, we do not attempt to probe the small-$x$ regime and will restrict our analyses to the region $x > 10^{-2}$.

To study the nuclear modification in the $e+A$ collisions in comparison with the $e+p$ collisions, we define the nuclear modification factor as the ratio of the OPEC in $e+A$ collisions to that in $e+p$ collisions:
\begin{align}
R_{eA}^{\mathrm{b.t.b.}}
\equiv
\frac{1}{A}
\frac{\dd{\Sigma^{\mathrm{b.t.b.}}_{eA}}}{\dd{x} \dd{Q^2} \dd{\tau}}
\Bigg{/}
\frac{\dd{\Sigma^{\mathrm{b.t.b.}}_{ep}}}{\dd{x} \dd{Q^2} \dd{\tau}}
, \label{eq:Ratio-b2bEC}
\end{align}
where $A$ is the atomic mass of the nuclei, and for gold nucleus we have $A = 197$.

\begin{figure}
\centering
\includegraphics[width = 0.46 \linewidth]{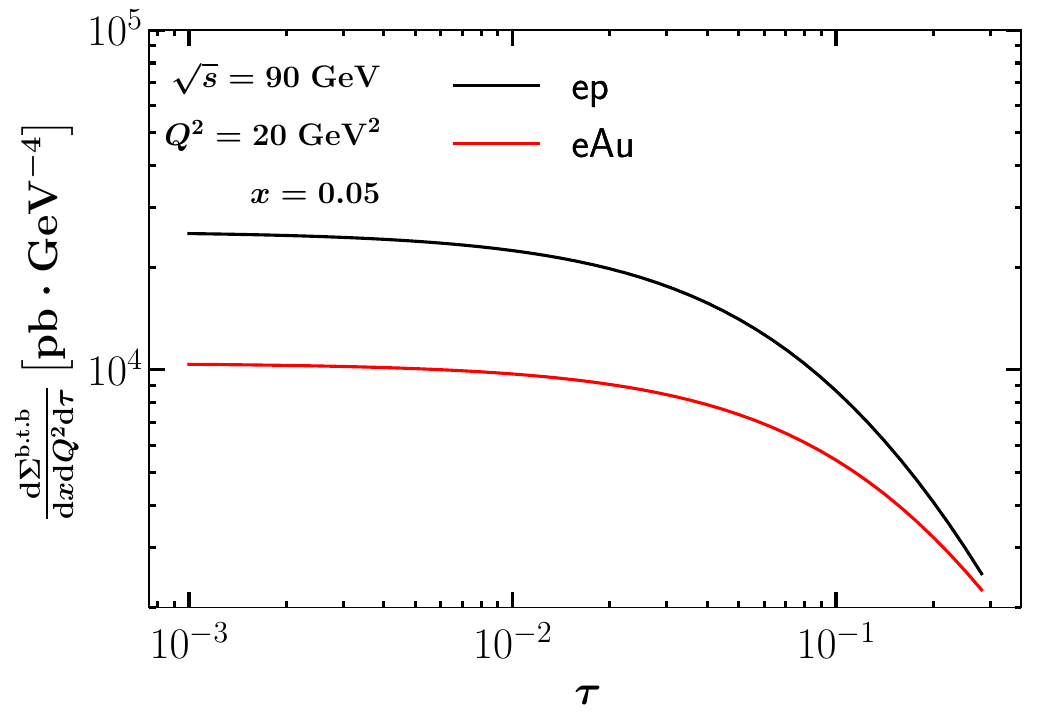}
\includegraphics[width = 0.46 \linewidth]{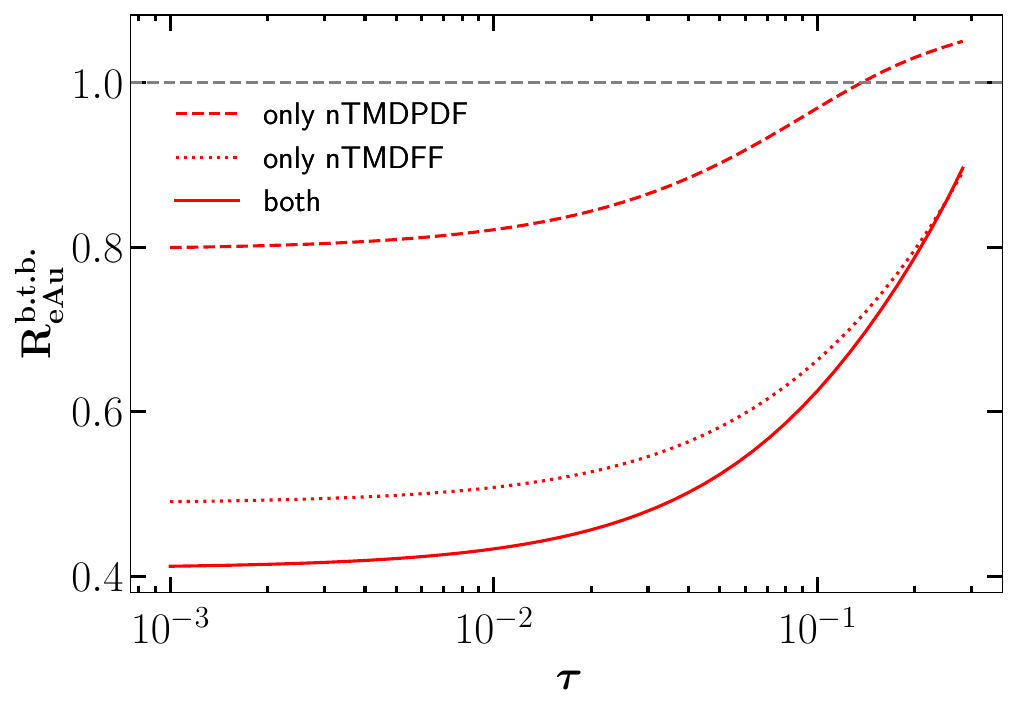}
\caption{
\textit{Left}: OPEC at the back-to-back region as a function of $\tau$.
We choose a center-of-mass energy of $\sqrt{s} = 90~\mathrm{GeV}$, photon's virtuality $Q^2 = 20~\mathrm{GeV}^2$ and $x = 0.05$.
The black and red solid curves represent the $e+p$ and $e+\mathrm{Au}$ results, respectively.
\textit{Right}: The corresponding nuclear modification factor $R_{eA}^{\mathrm{b.t.b.}}$ in \cref{eq:Ratio-b2bEC} as a function of $\tau$.
The red solid curve represents the complete calculation of the numerator in $R_{eA}^{\mathrm{b.t.b.}}$ using both nTMD PDFs and nTMD FFs.
In contrast, the red dashed line shows the $R_{eA}^{\mathrm{b.t.b.}}$ obtained using nTMD PDFs and \textit{vacuum} TMD FFs, whereas the dotted line is computed using the \textit{vacuum} TMD PDFs and nTMD FFs.
}
\label{fig:EEC_DIS_vs_tau}
\end{figure}

In the left panel of \cref{fig:EEC_DIS_vs_tau}, we plot the OPEC observable, $\frac{\dd{\Sigma^{\mathrm{b.t.b.}}}}{\dd{x} \dd{Q^2} \dd{\tau}}$, as a function of $\tau$ at $x = 0.05$.
The black solid curve represents the $e+p$ OPEC cross section, while the red solid curve corresponds to that of $e+\mathrm{Au}$.
The corresponding nuclear modification factor $R_{eA}^{\mathrm{b.t.b.}}$, defined in \cref{eq:Ratio-b2bEC}, is shown as the solid red curve in the right panel of \cref{fig:EEC_DIS_vs_tau}.
In the present kinematic setup, the nuclear suppression is about $0.33$ in the small-$\tau$ region and gradually weakens as $\tau$ increases.
To disentangle the contributions from initial- and final-state effects, we present two reference calculations: the red dashed curve is computed using nTMD PDFs together with vacuum TMD FFs, while the red dotted curve is computed using vacuum TMD PDFs together with nTMD FFs. By comparing the dashed, dotted and solid curves in the right panel of \cref{fig:EEC_DIS_vs_tau}, we conclude that the overall suppression comes from the combined effects of the nTMD PDFs and nTMD FFs, both of which can suppress the magnitude of the OPEC observable.

We further note that the dependence of $R_{eA}^{\mathrm{b.t.b.}}$ on the nTMD PDFs and nTMD FFs are both non-trivial in $\tau$.
This is expected because the $\tau$ dependence of the OPEC observable in \cref{eq:EECDIS_sub} gets convoluted between the TMD PDFs and TMD FFs (or EEC jet functions) via the Bessel function $J_0 \pqty{b \sqrt{\tau Q^2}}$.

\begin{figure}
\centering
\includegraphics[width = 0.46 \linewidth]{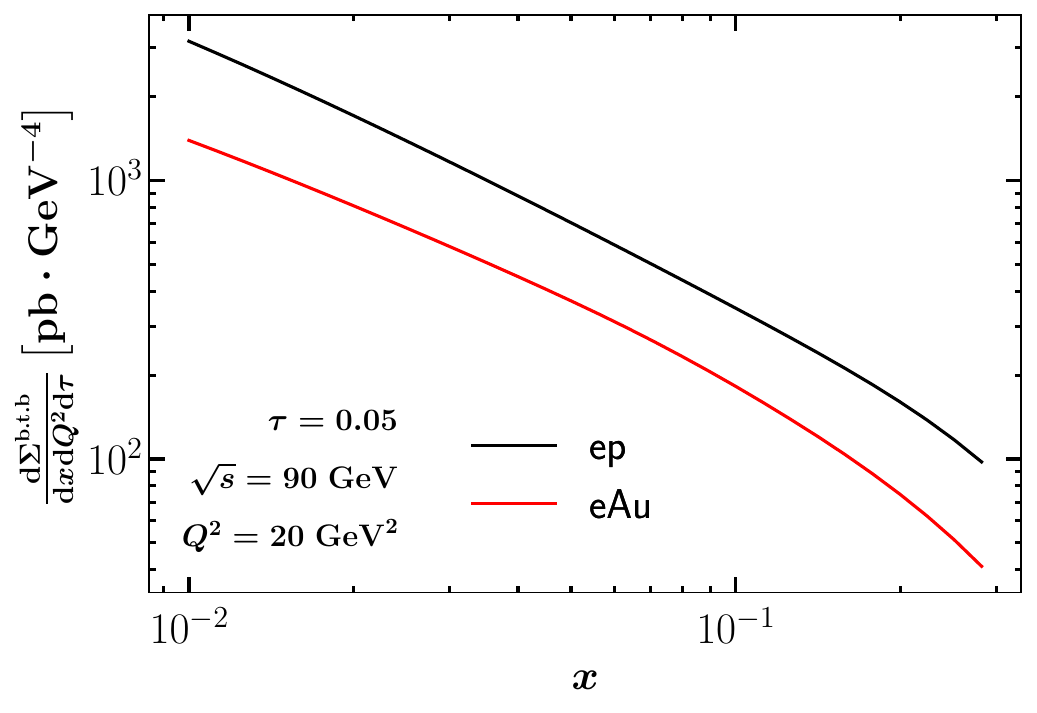}
\includegraphics[width = 0.46 \linewidth]{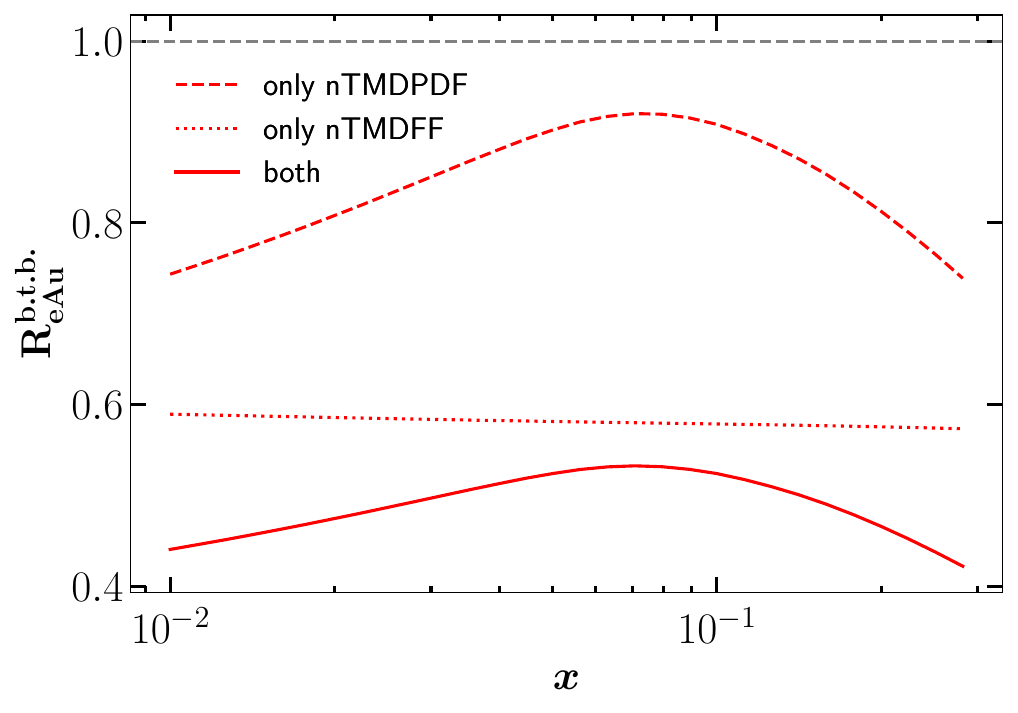}
\caption{
\textit{Left}: OPEC at the back-to-back region for $e+p$ and $e+\mathrm{Au}$ collisions as a function of Bjorken $x$.
\textit{Right}: The corresponding nuclear modification factor as a function of Bjorken $x$.
The meaning of dashed and dotted red lines in the right panel is identical to that in \cref{fig:EEC_DIS_vs_tau}.
}
\label{fig:EEC_DIS_vs_x}
\end{figure}

In the left panel of \cref{fig:EEC_DIS_vs_x}, we plot the OPEC, $\frac{\dd{\Sigma^{\mathrm{b.t.b.}}}}{\dd{x} \dd{Q^2} \dd{\tau}}$, as a function of $x$ at $\tau = 0.05$.
We observe that the correlator decreases with increasing $x$, and this behavior is primarily driven by the $x$-dependent overall factor in the factorization formula \labelcref{eq:EECDIS_sub}.
The corresponding nuclear modification factor $R_{eA}^{\mathrm{b.t.b.}}$ is shown in the right panel.
The modification induced by the nTMD FFs is found to be largely insensitive to Bjorken $x$, as expected, while the turnover behavior observed in the solid and dashed curves reflects the $x$-dependence in the nTMD PDFs.


\section{One-point energy correlator at collinear limit}
\label{sec:collinear}

In this section, we study the in-jet OPEC that measures the energy correlation between the jet and the hadron inside the jet, at the limit where the hadron is almost collinear with the jet axis.
The in-jet OPEC is studied via the back-to-back electron-jet production in DIS.
The back-to-back configuration between the scattered electron and the jet generates a small transverse momentum imbalance between them, which is naturally applicable for TMD factorization.
In addition, at the back-to-back limit, the photon-quark scattering channel produces a quark-initiated jet that recoils against the outgoing electron in the transverse plane at leading order, while the photon-gluon fusion channel is power suppressed.
Consequently, this kinematic selection allows for a cleaner probe of quark TMD dynamics and the nuclear modification of transverse momentum broadening.
We will first set up the kinematics and then present a factorized formula for the in-jet OPEC within the TMD framework.
Based on the formula, we will present numerical results relevant to future EIC phenomenology, focusing on the modifications induced by the presence of cold nuclear matter.

\subsection{Kinematics}
\label{sec:leptonjet-kinematics}

\begin{figure}
\centering
\includegraphics[width = 0.8 \textwidth]{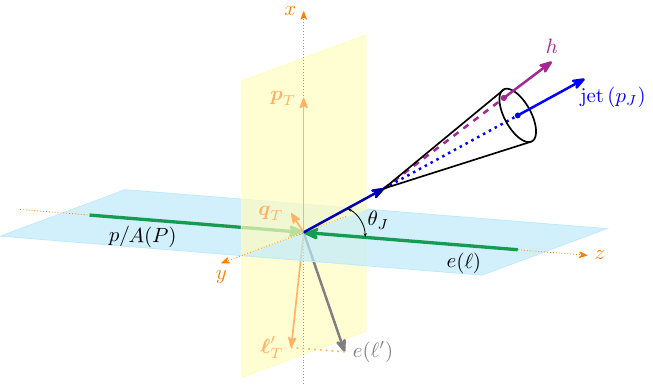}
\caption{
Back-to-back electron and jet production in unpolarized electron-proton or electron-nucleus collisions.
A hadron within the jet is measured.
The jet axis and the beam direction define the $xz$-plane.
A detailed illustration of the jet cone and hadron kinematics is given in \cref{fig:cone}.
}
\label{fig:qT}
\end{figure}

\begin{figure}
\centering
\includegraphics[width = 0.7 \textwidth]{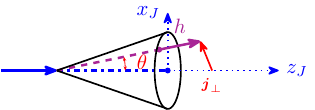}
\caption{
A hadron inside a jet characterized by its transverse momentum $\bm{j}_{\perp}$ and polar angle $\theta$, both measured relative to the jet axis $z_J$.}
\label{fig:cone}
\end{figure}

We consider the back-to-back production of an electron and a jet from the unpolarized electron-proton/nucleus collision:
\begin{align}
e \pqty{\ell} + p/A \pqty{P}
\to
e \pqty{\ell'}
+
\mathrm{jet} \pqty{{p}_J} \bqty{h \pqty{z_h, \bm{j}_{\perp}}}
+
X
,
\end{align}
where an electron with momentum $\ell$ (moving along the $-z$ direction) scatters with an unpolarized proton/nucleus with momentum $P$ (moving along the $+z$ direction), producing an electron with momentum $\ell'$ and a jet with momentum $p_J$.
In addition, a hadron is measured inside the jet.
The variable $z_h$ is the longitudinal momentum fraction of the jet carried by the hadron $h$, and $\bm{j}_{\perp}$ is the hadron's transverse momentum with respect to the jet axis.
The scattering process is illustrated in \cref{fig:qT}, while the hadron kinematics inside the jet are depicted in \cref{fig:cone}.
Note the subscript ``$T$'' denotes the transverse momenta measured with respect to the incoming beam direction, while the subscript ``$\perp$'' denotes the transverse momenta relative to the jet axis.

We study the scattering process using the center-of-mass frame of the $e+p/A$ collision.
Under the high-energy limit, the incoming momenta can be written as:
\begin{align}
P^{\mu}
=
\frac{\sqrt{s}}{2} \pqty{1,0,0,1}
, \quad
\ell^{\mu}
=
\frac{\sqrt{s}}{2} \pqty{1,0,0,-1}
, \label{eq:light-cone}
\end{align}
where $s = \pqty{P + \ell}^2$ is the center-of-mass energy squared.
We define the virtuality of the exchanged photon $Q^2 \equiv -q^2= -\pqty{\ell-\ell'}^2$, and the event inelasticity $y \equiv Q^2/\pqty{x s}$ with $x \equiv Q^2/ \pqty{2P \cdot q}$ being the standard Bjorken $x$ variable.
The rapidity of the jet is defined as $\eta_J \equiv -\ln(\tan(\theta_J/2))$ with $\theta_J$ being the jet's polar angle, and the jet's four-momentum reads:
\begin{align}
p_J^{\mu}
=
E_J \pqty{1, \sin{\theta_J}, 0, \cos{\theta_J}}
, \label{eq:jetmom}
\end{align}
where $E_J$ is the jet energy and we have put the jet momentum within the $xz$-plane (see \cref{fig:qT}).
Finally, we denote the jet radius by $R$.

We consider the kinematic region where the scattered electron and the jet are produced almost back-to-back in the $xy$-plane, resulting in a transverse momentum imbalance $\bm{q}_T$ that has a much smaller magnitude than that of the average transverse momentum $\bm{p}_T$.
Here, $\bm{q}_T$ and $\bm{p}_T$ are defined as:
\begin{align}
\bm{q}_T
\equiv
\bm{p}_{J,T} + \bm{\ell}'_T
, \quad
\bm{p}_T
\equiv
\pqty{\bm{p}_{J,T} - \bm{\ell'}_T}/2
, \label{eq:qTandpT}
\end{align}
where $\bm{p}_{J,T}$ and $\bm{\ell}'_T$ are the transverse momenta of the jet and the outgoing electron, respectively.


\subsection{Factorization formalism}
\label{ss.factorization-collinear-OPEC}

Working in the one-photon exchange approximation and neglecting the electron mass, the differential cross section of the back-to-back electron-jet production with an unpolarized hadron measured inside the jet is given by \cite{Kang:2021ffh}:
\begin{align}
\frac{\dd{\sigma_{ e + \mathrm{jet}\pqty{h} }}}{\dd{p_T^2} \dd{\eta_J} \dd{q_T^2} \dd{z_h} \dd{j_{\perp}^2}}
=
\frac{(2\pi)^2}{4} F_{UU,U}
, \label{eq:unpjethintang}
\end{align}
where the azimuthal angles of $\bm{q}_T$ and $\bm{j}_{\perp}$ have been integrated out.
The structure function $F_{UU,U}$ reads \cite{Buffing:2018ggv, Arratia:2020nxw, Kang:2020xez, Kang:2021ffh}%
\footnote{{In principle, the so-called non-global logarithms (NGLs) contribute to the jet cross section. Their full treatment leads to multi-Wilson-line structures and complicated matrix-valued evolution~\cite{Becher:2016mmh}. However, using the prescription of Ref.~\cite{Dasgupta:2001sh} and following our earlier work~\cite{Arratia:2020nxw}, we find their contribution to be relatively small ($\lesssim 4\%$ in our kinematic regime), and we therefore neglect them in the present analysis.}}:
\begin{align}
F_{UU,U}
& =
\hat{\sigma}_0
H (Q,\mu)
\sum_q
e_q^2 \mathcal{D}_{h/q} \pqty{z_h, j_{\perp}, \mu, \zeta_J}
\nonumber \\
& \quad \times
\int \frac{b \dd{b}}{2 \pi}
J_0 \pqty{q_T b}
x f_q^{\pqty{u}} \pqty{x,b, \mu,\zeta}
\overline{S}_{\mathrm{global}}^{\pqty{u}} \pqty{b,\mu}
\overline{S}_{\mathrm{cs}} \pqty{b,R,\mu}
, \label{eq:FUUUbefore}
\end{align}
where we include renormalization scale $\mu$ and Collins-Soper parameter $\zeta_J$ for the TMD fragmenting jet functions (TMD FJFs) $\mathcal{D}_{h/q} \pqty{z_h, j_{\perp}^2, \mu, \zeta_J}$ and $j_{\bot} \equiv \abs{\bm{j}_{\bot}}$.
As demonstrated in \cref{ss.RG-consistency-at-collinear-limit}, we choose $\sqrt{\zeta_J} = p_T R$ complying with the requirement from the RG consistency.
In addition, at the Born-level of the unpolarized scattering $e \pqty{\ell} + q \pqty{xP} \to e \pqty{\ell'} + q \pqty{p_J}$, the momentum fraction carried by the initial quark with respect to its parent proton is equal to the Bjorken $x$.
The cross section for this scattering process is given by:
\begin{align}
\hat{\sigma}_0
=
\frac{\alpha_{\mathrm{em}}^2}{s Q^2}
\frac{2 \pqty{\hat{u}^2 + \hat{s}^2}}{\hat{t}^2} =\frac{\sigma_0}{\pi y s}
, \label{eq:sig0}
\end{align}
where $\sigma_0$ was defined in \cref{eq:hard-func}, $\hat{s}$, $\hat{t}$ and $\hat{u}$ are the partonic Mandelstam variables that are defined as:
\begin{align}
\hat{s}
\equiv
\pqty{xP + \ell}^2
, \quad
\hat{t}
\equiv
\pqty{\ell - \ell'}^2
, \quad
\hat{u}
\equiv
\pqty{xP - \ell'}^2
.
\end{align}
Additionally, $H(Q, \mu)$ is the hard function, and its renormalized expression at NLO is given in \cref{eq:hard-func-NLO}.
The one-loop expression of the azimuthal angle averaged ``unsubtracted'' global soft function is \cite{Hornig:2017pud, Kang:2020xez}:
\begin{align}
\overline{S}_{\mathrm{global}}^{\pqty{u}} \pqty{b, \mu, \nu}
& =
1
-
\frac{\alpha_s C_F}{2\pi}
\bigg[
\pqty{\frac{2}{\eta} + \ln(\frac{\nu^2}{\mu^2}) - 2\eta_J}
\pqty{\frac{1}{\epsilon} + \ln(\frac{\mu^2}{\mu_b^2})}
\nonumber \\
& \qquad \qquad \qquad \quad -
\frac{2}{\epsilon^2}
-
\frac{1}{\epsilon} \ln(\frac{\mu^2}{\mu_b^2})
+
\frac{\pi^2}{6}
\bigg] ,
\end{align}
and it describes the soft radiation throughout the full phase space without resolving the jet cone.
On the other hand, the collinear-soft function $\overline{S}_{\mathrm{cs}}$ is sensitive only to the soft radiation around the jet direction and resolves the jet cone.
Its expression is given by \cite{Hornig:2017pud, Kang:2020xez}:
\begin{align}
\overline{S}_{\mathrm{cs}} \pqty{b, R, \mu}
=
1
-
\frac{\alpha_s C_F}{2\pi}
\bqty{
\frac{1}{\epsilon^2}
+
\frac{1}{\epsilon} \ln(\frac{\mu^2}{\mu_b^2 R^2})
+
\frac{1}{2}\ln[2](\frac{\mu^2}{\mu_b^2 R^2})
-
\frac{\pi^2}{12}
} ,
\end{align}

Following \cref{e.subtracted-TMD-PDFs-definition}, we define the ``subtracted'' TMD PDFs $f_q \pqty{x, b, \mu, \zeta}$.
Accordingly, we also introduced the ``subtracted'' global soft function that is free from rapidity divergence:
\begin{align}
\overline{S}_{\mathrm{global}} \pqty{b, \mu}
=
\frac{\overline{S}_{\mathrm{global}}^{\pqty{u}} \pqty{b, \mu, \nu}}{\sqrt{S_{n \overline{n}} \pqty{b, \mu, \nu}}}
,
\end{align}
where $S_{n \overline{n}} \pqty{b, \mu, \nu}$ is defined in \cref{eq:Snnbar}.

In the kinematic region $j_{\perp} \ll p_T R$, the unpolarized TMD FJFs $\mathcal{D}_{h/q} \pqty{z_h, j_{\perp}, \mu, \zeta_J}$ can be further factorized in terms of the corresponding unpolarized TMD FFs and an in-jet soft function as \cite{Kang:2017glf, Kang:2020xyq}:
\begin{align}
\mathcal{D}_{h/q} \pqty{z_h, j_{\perp}, \mu, \zeta_J}
& =
\int \dd[2]{\bm{k}_{\perp}} \dd[2]{\bm{\lambda}_{\perp}}
\delta^2 \pqty{z_h \bm{\lambda}_{\perp} + \bm{k}_{\perp} - \bm{j}_{\perp}}
D_{h/q}^{(u)} \pqty{z_h, k_{\perp}, \mu, \frac{\zeta'}{\nu^2}}
S_q \pqty{\lambda_{\perp}, \mu, \nu \mathcal{R}}
\nonumber \\
& =
\int \frac{b \dd{b}}{2\pi}
J_0 \pqty{\frac{j_{\perp} b}{z_h}}
D_{h/q}^{(u)} \pqty{z_h, b, \mu, \frac{\zeta'}{\nu^2}}
S_q \pqty{b, \mu, \nu \mathcal{R}}
, \label{unp_JFF_FF}
\end{align}
where $\zeta'$ is the Collins-Soper parameter for the TMD FFs and it satisfies $\zeta_J = \zeta'\mathcal{R}^2/4$ with $\mathcal{R} \equiv R/\cosh{\eta_J}$.
The Fourier transformation of the ``unsubtracted'' TMD FFs $D_{h/q}^{(u)}$ is defined in \cref{eq:FF-FT}.
Meanwhile, the in-jet soft function $S_q \pqty{b, \mu, \nu \mathcal{R}}$ in $b$-space is given by \cite{Kang:2017glf, Buffing:2018ggv, Kang:2021ffh}:
\begin{align}
S_q \pqty{b, \mu, \nu \mathcal{R}}
=
1
& -
\frac{\alpha_s C_F}{4\pi}
\bigg[
2 \pqty{\frac{2}{\eta} + \ln(\frac{\nu^2\mathcal{R}^2}{4\mu^2})}
\pqty{\frac{1}{\epsilon} + \ln(\frac{\mu^2}{\mu_b^2})}
\nonumber \\
& \qquad \qquad \quad +
\ln[2](\frac{\mu^2}{\mu_b^2})
-
\frac{2}{\epsilon^2}
+
\frac{\pi^2}{6}
\bigg]
, \label{eq:S_q}
\end{align}
where we have applied the narrow jet approximation $\tan(\mathcal{R}/2) \approx \mathcal{R}/2$.
It is instructive to realize that $S_q \pqty{b, \mu, \nu \mathcal{R}} = \sqrt{S_{n \overline{n}} \pqty{b, \mu, \nu}}|_{\nu \to \nu \mathcal{R}/2}$ at the NLO \cite{Kang:2017glf, Kang:2023elg}, where $S_{n  \overline{n}}$ is the standard soft function given in \cref{eq:Snnbar}.
One can therefore define the ``subtracted'' in-jet TMD FFs $D_{h/q}$ as:
\begin{align}
D_{h/q} \pqty{z_h, b, \mu, \frac{\zeta' \mathcal{R}^2}{4}}
& \equiv
D_{h/q}^{(u)} \pqty{z_h, b, \mu, \frac{\zeta'}{\nu^2}}
S_q \pqty{b, \mu, \nu \mathcal{R}}
\nonumber \\
& =
D_{h/q}^{(u)} \pqty{z_h, b, \mu, \frac{\zeta'}{\nu^2}}
\sqrt{S_{n  \overline{n}} \pqty{b, \mu, \frac{\nu \mathcal{R}}{2}}}
. \label{eq:Dunpb}
\end{align}
Finally, we obtain the following relation between TMD FJFs and the ``subtracted'' in-jet TMD FFs:
\begin{align}
\mathcal{D}_{h/q} \pqty{z_h, j_{\perp}, \mu, \zeta_J}
=
\int \frac{b \dd{b}}{2\pi}
J_0 \pqty{\frac{j_\perp b}{z_h}}
D_{h/q} \pqty{z_h, b, \mu, \zeta_J}
=
D_{h/q} \pqty{z_h, j_{\perp}, \mu, \zeta_J}
, \label{eq:TMDFFandTMDFJF}
\end{align}
meaning that, at NLO, the TMD FJFs are equal to the ``subtracted'' in-jet TMD FFs at the scale $\zeta_J$.
At this point, we can rewrite \cref{eq:FUUUbefore} so that the TMD distributions are free from rapidity divergence:
\begin{align}
F_{UU,U}
& =
\hat{\sigma}_0
H \pqty{Q, \mu}
\sum_q
e_q^2 
\int \frac{b' \dd{b'}}{2\pi}
J_0 \pqty{\frac{j_\perp b'}{z_h}}
D_{h/q} \pqty{z_h, b', \mu, \zeta_J}
\nonumber \\
& \quad \times
\int \frac{b \dd{b}}{2\pi}
J_0 \pqty{q_T b}
x f_q \pqty{x, b, \mu, \zeta}
\overline{S}_{\mathrm{global}} \pqty{b, \mu}
\overline{S}_{\mathrm{cs}} \pqty{b, R, \mu}
. \label{eq:FUUUafter}
\end{align}
Here the ``unsubtracted'' TMD PDFs have also been substituted by the ``subtracted'' TMD PDFs \labelcref{e.subtracted-TMD-PDFs-definition}.

For the purpose of studying the OPEC inside jet, instead of measuring $j_{\perp}$, one measures the angle between the hadron's momentum and the jet axis, denoted as $\theta$, as shown in \cref{fig:cone}.
At the small-$j_{\perp}$ limit, \textit{i.e.}, $j_{\perp}\ll z_h E_J$, we have $ \theta \approx \frac{j_{\perp}}{z_h p_T \cosh \eta_J}$.
For the back-to-back electron-jet production with unpolarized hadrons observed inside jet, we define the OPEC inside the jet as:
\begin{align*}
\frac{\dd{\Sigma^{\mathrm{coll.}}}}{\dd{p_T^2} \dd{\eta_J} \dd{q_T^2} \dd{\theta^2}}
=
\sum_h
\int \dd{z_h} z_h
\int \dd{j_{\perp}^2}
\delta \pqty{\theta^2 - \frac{j_{\perp}^2}{z_h^2 p_T^2 \cosh[2](\eta_J)}}
\frac{\dd{\sigma_{ e + \mathrm{jet} \pqty{h} }}}{\dd{p_T^2} \dd{\eta_J} \dd{q_T^2} \dd{z_h} \dd{j_{\perp}^2}}
.
\end{align*}
Plugging \cref{eq:unpjethintang,eq:FUUUafter} into the above equation, we get:
\begin{align}
\frac{\dd{\Sigma^{\mathrm{coll.}}}}{\dd{p_T^2} \dd{\eta_J} \dd{q_T^2} \dd{\theta^2}}
& =
\pqty{\pi p_T \cosh{\eta_J}}^2
\hat{\sigma}_0
H \pqty{Q, \mu}
\sum_q
e_q^2
\int \frac{b' \dd{b'}}{2 \pi}
J_0 \pqty{\theta\, b'\, p_T\cosh \eta_J }
J_q \pqty{b', \mu, \zeta_J}
\nonumber \\
& \quad \times
\int \frac{b \dd{b}}{2 \pi}
J_0 \pqty{q_T b}
x f_q \pqty{x, b, \mu, \zeta}
\overline{S}_{\mathrm{global}} \pqty{b, \mu}
\overline{S}_{\mathrm{cs}} \pqty{b, R, \mu}
, \label{eq:dtheta2}
\end{align}
where $J_q$ is the EEC jet function defined in \cref{eq:EECjetfunction}.

Experimentally, the measurement is typically performed on the $\eta\phi$-plane.
Accordingly, we define the ``measured angle'' between hadron and jet axis as $R_L \equiv \sqrt{\pqty{\Delta\eta}^2 + \pqty{\Delta\phi}^2}$, where $\Delta\eta$ and $\Delta\phi$ are the pseudo-rapidity and azimuthal angle difference between the observed hadron and the jet axis, respectively.
For small $R_L$, one can find the relation:
\begin{align}
\theta
=
\frac{R_L}{\cosh{\eta_J}}
\,.
\end{align}
Finally, we obtain the factorization formula for the in-jet OPEC with the back-to-back lepton-jet production in lepton-hadron collisions:
\begin{align}
\frac{\dd{\Sigma^{\mathrm{coll.}}}}{\dd{p_T^2} \dd{\eta_J} \dd{q_T^2} \dd{R_L}}
& =
2 \pi^2 R_L p_T^2
\hat{\sigma}_0
H \pqty{Q, \mu}
\sum_q
e_q^2
\int \frac{b' \dd{b'}}{2 \pi}
J_0 \pqty{R_L p_T b'}
J_q \pqty{b', \mu, \zeta_J}
\nonumber \\
& \quad \times
\int \frac{b \dd{b}}{2\pi}
J_0 \pqty{q_T b}
x f_q \pqty{x, b, \mu, \zeta}
\overline{S}_{\mathrm{global}} \pqty{b, \mu}
\overline{S}_{\mathrm{cs}} \pqty{b, R, \mu}
. \label{eq:injetEC-RL}
\end{align}
The treatment of TMD PDFs $f_q$ and the EEC jet function $J_q$ have been outlined in \cref{sec:Quark-Distribuition-function} and \cref{sec:EEC-jet-function}, respectively. We also note that here the angle-dependent term $J_0 \pqty{R_L p_T b'}$ only convolutes with the EEC jet function, hence we do not expect the TMD PDFs to influence the angular behavior of the OPEC in this case.
{One might wonder that the framework accounts for nuclear effects on hadrons inside the jet through the nTMD FFs, but does not seem to include potential modifications to the jet production rate itself in the nuclear medium.
It is worth noting that our framework follows the same philosophy as the standard nuclear collinear PDF approach \cite{Eskola:2016oht}, where all cold nuclear matter effects (including those on jet production) are absorbed into the nTMD PDFs, without modifying any other part of the formalism \cite{Alrashed:2021csd}.
In other words, the nTMD PDFs encode the nuclear modification of jet production, while the nTMD FFs encode the modification of the hadron distribution inside the jet.}
\subsection{Phenomenology}

In this section, we present numerical predictions for the in-jet OPEC at collinear limit with the jet produced in lepton-hadron collision, based on the factorization formalism established in \cref{eq:injetEC-RL}.
Our focus will be on the nuclear modification of the OPEC.
To this end, we compare the results for $e+\mathrm{Au}$ collisions to those for $e+p$ collisions.

In our numerical analysis, we consider $e+p$ and $e+A$ collisions at a center-of-mass energy of $\sqrt{s} = 90~\mathrm{GeV}$.
The transverse momentum imbalance is fixed to $q_T = 0.5~\mathrm{GeV}$, and the jet transverse momentum is set to $p_T = 10~\mathrm{GeV}$.
We vary the jet rapidity $\eta_J$ among $-1$, 0 and 1.
Using:
\begin{equation}
x
=
\frac{p_T \sqrt{s} e^{\eta_J}}{s - p_T \sqrt{s} e^{-\eta_J}}
,
\end{equation}
we find that the corresponding Bjorken $x$ values probed by the above $\eta_J$ values are approximately 0.045, 0.125 and 0.267, respectively.
The jet radius is fixed to $R = 1$.

\begin{figure}[t]
\centering
\includegraphics[width = 0.95 \linewidth]{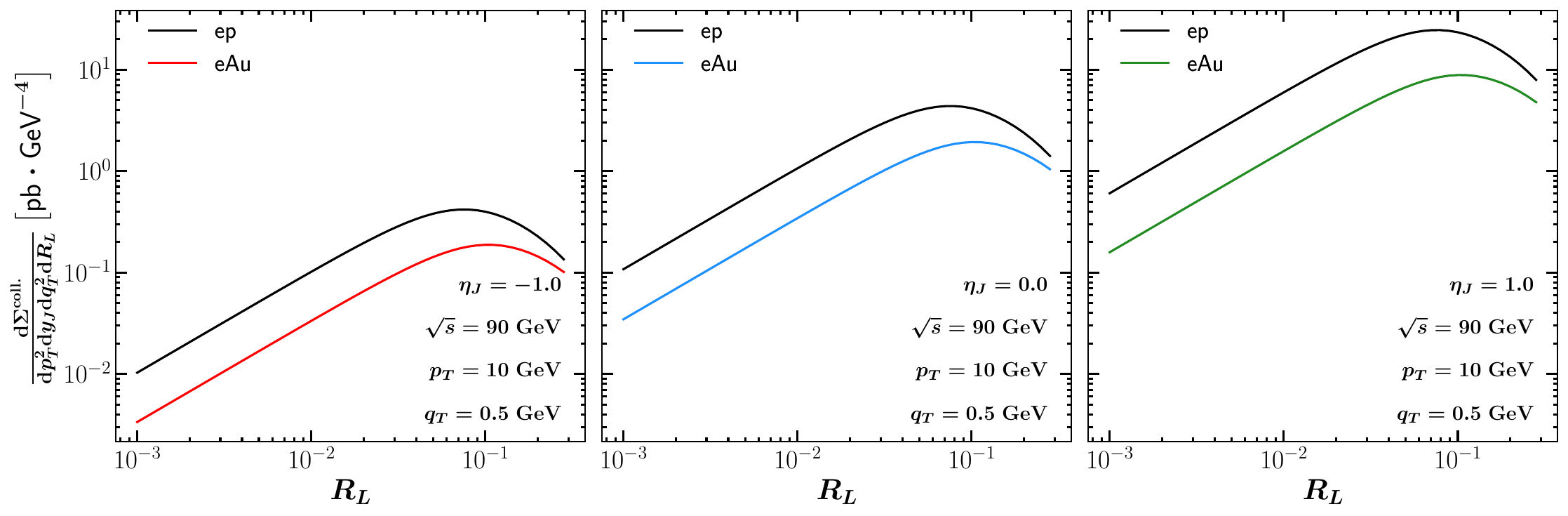}
\caption{
OPEC inside jet for $e+p$ and $e+\mathrm{Au}$ collisions as a function of $R_L$.
We choose a center-of-mass energy of $\sqrt{s} = 90~\mathrm{GeV}$.
Additionally, the transverse momentum imbalance $q_T$ is set to 0.5 GeV, while the jet transverse momentum is fixed at $p_T = 10~\mathrm{GeV}$.
Finally, the jet radius $R$ is 1.
The left, central and right panels correspond to jet rapidities $\eta_J = -1$, 0 and 1, respectively.
In each panel, the solid black curves represent the $e+p$ results, while the solid colored curves correspond to the $e+\mathrm{Au}$ results.}
\label{fig:inet_EC}
\end{figure}

In each panel of \cref{fig:inet_EC}, we present the OPEC, $\dd{\Sigma^{\mathrm{coll.}}} / \pqty{\dd{p_T^2} \dd{\eta_J} \dd{q_T^2} \dd{R_L}}$, as a function of $R_L$ for the three rapidity values.
The solid black curves correspond to the $e+p$ results, while the solid colored curves depict the results from $e+\mathrm{Au}$ collision.
Their qualitative features closely resemble those observed for the in-jet EEC in \cite{Komiske:2022enw}.
Analogous to the case of the EEC, the opening angle $R_L$ can be used as a proxy for the virtuality scale that governs the in-jet shower evolution.
The jet evolution time is parametrically encoded in the angular scale of the correlator as $t \propto 1/(p_T R_L^2)$.
We find that the $e+p$ OPEC observables peak at $R_L^{\mathrm{peak}} \sim 0.08$.
For $R_L < R_L^{\mathrm{peak}}$, this small-angle region corresponds to a late time stage of the jet showering. The OPEC observable lies in the nonperturbative region and exhibits a characteristic linear rise as $R_L$ increases, which implies that the energy is uniformly distributed and can be explained by the fact that free hadrons carry it \cite{Komiske:2022enw}. This region is typically referred to as the free-hadron or non-perturbative region.
As $R_L$ grows toward the peak, the behavior reflects the transition from the small-angle non-perturbative region to the large-angle perturbative region.
Beyond the peak point, the OPEC begins to decrease with increasing $R_L$, eventually approaching the large-angle regime ($R_L \gtrsim 0.3$) where the TMD factorization framework is no longer applicable and the dynamics are instead governed by fixed-order calculations within collinear QCD.
This region, however, is beyond the scope of this work.
It would be interesting to further quantify the relationship between the EEC and OPEC within jets, and we leave such an investigation for future work.

\begin{figure}[t]
\centering
\includegraphics[width = 0.95 \linewidth]{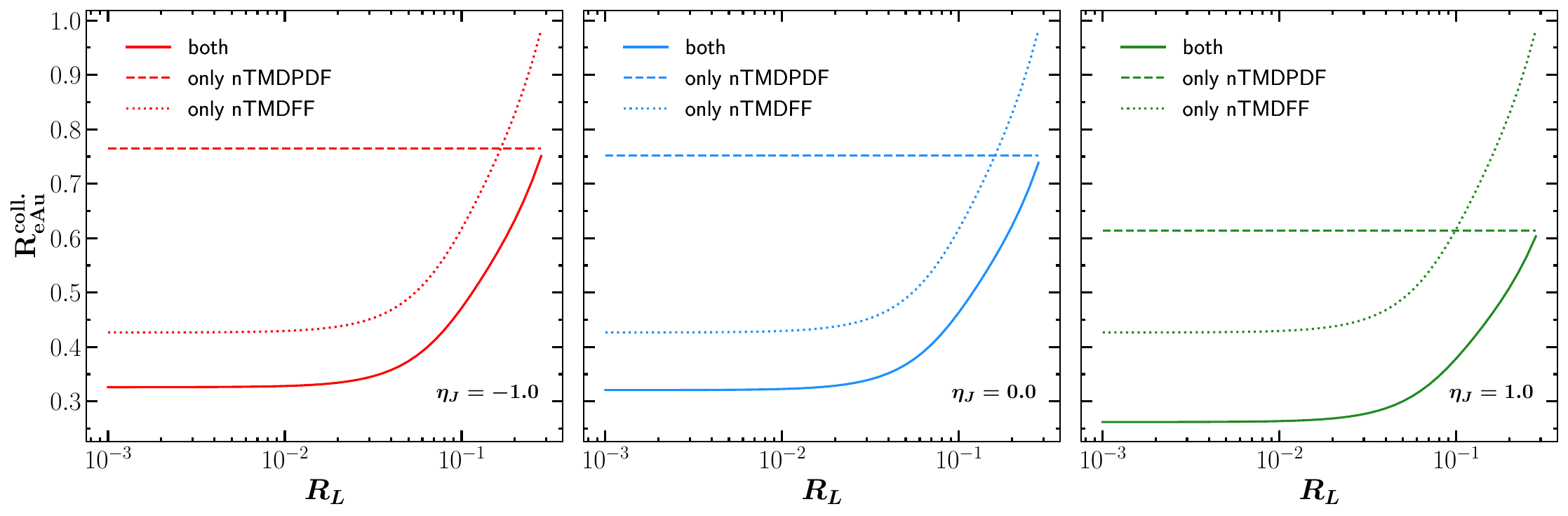}
\caption{
Nuclear modification factor $R_{eA}^{\mathrm{coll.}}$ for $e+\mathrm{Au}$ collisions at jet rapidities $\eta_J = -1$, 0 and 1, respectively.
The solid curves represent the ratio of $e+\mathrm{Au}$ results to those of $e+p$, as defined in \cref{e.R_eA-collinear-OPEC}.
The dashed lines represent the same ratio as the solid curve, except that the numerator in \cref{e.R_eA-collinear-OPEC} is computed with \textit{vacuum} TMD FFs and nTMD PDFs.
The dotted lines, on the other hand, are obtained with the numerator calculated with \textit{vacuum} TMD PDFs and nTMD FFs.
}
\label{fig:inet_EC_ratio}
\end{figure}

To quantify the nuclear effects in $e+A$ collision relative to $e+p$ collision, we define the nuclear modification factor as the ratio of the OPEC in $e+A$ to that in $e+p$:
\begin{align}
R_{eA}^{\mathrm{coll.}}
\equiv
\frac{1}{A} 
\frac{\dd{\Sigma^{\mathrm{coll.}}_{eA}}}
{\dd{p_T^2} \dd{\eta_J} \dd{q_T^2} \dd{R_L}}
\Bigg/
\frac{\dd{\Sigma^{\mathrm{coll.}}_{ep}}}
{\dd{p_T^2} \dd{\eta_J} \dd{q_T^2} \dd{R_L}}
, \label{e.R_eA-collinear-OPEC}
\end{align}
where $A$ is the atomic mass number of the nucleus, and for gold we have $A = 197$.
The nuclear modification factors are shown in \cref{fig:inet_EC_ratio} as a function of $R_L$, where the solid curves represent the complete results, \textit{i.e.}, the numerator in \cref{e.R_eA-collinear-OPEC} is computed with both nTMD PDFs and nTMD FFs.
For the kinematic region under consideration, we find that the nuclear modification factors vary between 0.25 and 0.35 in the small-$R_L$ region.
The suppression weakens as the angle $R_L$ increases, which is expected because the nuclear effect tends to broaden the transverse momentum distribution.

To disentangle different sources of nuclear effects, two additional reference calculations are shown in \cref{fig:inet_EC_ratio} as dashed and dotted curves.
In the dashed curves, the initial-state nuclear effects are isolated by employing the vacuum TMD FFs in the numerator of \cref{e.R_eA-collinear-OPEC}.
They indicate that the nuclear modification arising from the nTMD PDFs is essentially independent of $R_L$.
Conversely, by substituting the nTMD PDFs with their vacuum counterpart, we isolate the final-state nuclear effects.
The resulting nuclear modification factors (dotted curves) follow a similar trend as the complete results (solid curves), indicating that final-state nuclear effects dominate the angular dependence of $R_{eA}^{\mathrm{coll.}}$.
These observations are consistent with the factorization formula we established in \cref{eq:injetEC-RL}.
Physically, the angular dependence of the in-jet OPEC is affected only by in-jet soft radiation, \textit{i.e.}, the correlation among final-state particles. Therefore, we do not expect the initial-state effect from the nTMD PDFs to affect the angular distribution inside the jet.

{In the future, it would be interesting to study the uncertainties 
of the nTMD PDFs and nTMD FFs, along with the perturbative 
uncertainties estimated via scale variations. We note that for 
the nuclear modification factors, a large part of the scale 
dependence is expected to cancel in the ratio, making them more 
robust observables for the extraction of nuclear effects. A 
dedicated study of these uncertainties would further clarify 
how nuclear effects can be quantitatively extracted from future 
experiments.}

\section{Conclusion}
\label{sec:conclusion}

In this work, we have explored two limits of the OPEC observables that can be measured in the future EIC.
We first studied the back-to-back limit of OPEC in DIS within the kinematic region where the opening angle $\theta_h$ between the final hadrons and the initial proton beam is close to $\pi$.
We provided a factorization formula within the TMD framework, based on which we presented numerical results for this observable in $e+p$ and $e+\mathrm{Au}$ collisions.
We found that the OPEC cross section decreases as $\tau$ increases.
In addition, we studied the nuclear modification with $e+\mathrm{Au}$ collisions.
We found that the nuclear modification factor $R_{eA}^{\mathrm{b.t.b.}}$ is approximately $0.4$ in the small $\tau$ region and grows as $\tau$ increases.
As expected, the nuclear medium broadens the transverse momentum distribution and smears the OPEC cross section to the large $\tau$ region.
In addition, we investigated the nuclear modification as a function of Bjorken $x$, and our findings showed that the nuclear modification of the OPEC can also provide valuable insights into the nuclear modification of the collinear PDFs.

Secondly, we studied the collinear limit of the OPEC observable inside the jet, which is measured in the back-to-back lepton-jet production.
We focused on the collinear limit where the angle $R_L$ between the hadron and jet axis is close to 0.
A factorization formula was provided within the TMD framework, and numerical results were presented for $e+p$ and $e+\mathrm{Au}$ collisions.
The in-jet OPEC increases as $R_L$ increases, but peaks at $R_L \sim 0.08$.
In $e+\mathrm{Au}$ collisions, the nuclear modification factor $R_{eA}^{\mathrm{coll.}}$ varies from 0.25 to 0.35 in the small-$R_L$ region, depending on the jet rapidity $\eta_J$.
As $R_L$ further increases, $R_{eA}^{\mathrm{coll.}}$ starts to grow, similar to the case of OPEC at the back-to-back limit.
Again, this is driven by the broadening of the transverse momentum distribution from the nuclear medium.
We further found that the angular behavior of $R_{eA}^{\mathrm{coll.}}$ is independent of the initial-state nuclear effects, which is expected as the $R_{eA}^{\mathrm{coll.}}$ of OPEC at the collinear limit probes the final-state transverse momentum broadening.

In summary, exploring the OPEC in back-to-back and collinear limits in $e+p$ and $e+A$ collisions presents a fertile ground for studying TMD physics and cold nuclear effects.
We anticipate the insights obtained from OPEC will play a pivotal role in studying cold nuclear matter effects at the upcoming EIC.

\section*{Acknowledgements}

We thank Berndt M\"{u}ller and Wouter Waalewijn for helpful discussions. Y.F. is supported by the grant DE-FG02-05ER41367 from the U.S. Department of Energy, Office of Science, Nuclear Physics.
Z.K. and J.P. are supported by the National Science Foundation under grant No.~PHY-2515057. This work is also supported by the U.S. Department of Energy, Office of Science, Office of Nuclear Physics, within the framework of the Saturated Glue (SURGE) Topical Theory Collaboration. 
Y.Z. is supported by the European Union ``Next Generation EU'' program through the Italian PRIN 2022 grant No.~20225ZHA7W.

\appendix

\section{Renormalization group consistency of $\mu$-evolution}
\label{App:RGC}

In this section, we provide a prescription for determining the Collins–Soper scales used in the factorization formalism, based on the consistency conditions of the $\mu$-evolution for OPEC in both the back-to-back and collinear limits.

\subsection{OPEC at the back-to-back limit}
\label{ss.RG-consistency-at-b-to-b-limit}

At the back-to-back limit, the factorization formula for the OPEC observable is given in \cref{eq:EECDIS_sub}.  
The rapidity scales $\zeta$ and $\hat{\zeta}$ appearing in the TMD PDFs and the EEC jet function are constrained by the requirement that the factorized cross section is independent of the renormalization scale $\mu$.
This leads to the renormalization group (RG) consistency condition.

At one loop, the $\mu$-anomalous dimensions of the hard function \cite{Ellis:2010rwa}, the quark TMD PDFs and the EEC jet function are:
\begin{align}
\gamma_{\mu}^H
& =
-2 \frac{\alpha_s C_F}{\pi}
\pqty{\ln(\frac{\mu^2}{Q^2}) + \frac{3}{2}}
, \label{eq:ad-hadrd} \\
\gamma_{\mu}^q
& =
\frac{\alpha_s C_F}{\pi}
\pqty{\ln(\frac{\mu^2}{\zeta}) + \frac{3}{2}}
, \label{eq:ad-qTMDPDF} \\
\gamma_{\mu}^J
& =
\frac{\alpha_s C_F}{\pi}
\pqty{\ln(\frac{\mu^2}{\hat{\zeta}}) + \frac{3}{2}}
. \label{e.mu-evolution-J_q}
\end{align}
RG consistency of the cross-section requires:
\begin{align}
\gamma_{\mu}^H + \gamma_{\mu}^q + \gamma_{\mu}^J = 0
,
\end{align}
which implies the relation:
\begin{align}
\zeta \hat{\zeta}
=
Q^4
.
\end{align}
Hence we choose $\zeta = \hat{\zeta} = Q^2$.

\subsection{OPEC at the collinear limit}
\label{ss.RG-consistency-at-collinear-limit}

Before considering the OPEC in electron-jet production, we first examine the RG consistency for jet production where the internal structure of the jet is not measured, namely
\begin{align}
e \pqty{\ell} + p/A \pqty{P}
\to
e \pqty{\ell'}
+
\mathrm{jet} \pqty{{p}_J} 
+
X\,.
\end{align}
Here, the kinematics and notations are the same as in \cref{sec:collinear}.
The differential cross section for this process can be expressed within the TMD factorization formalism as \cite{Buffing:2018ggv, Arratia:2020nxw, Kang:2020xez, Kang:2021ffh}:
\begin{align}
\frac{\dd{\sigma_{e+\mathrm{jet}+X}}}{\dd{{p_T^2}} \dd{\eta_J} \dd[2]{\bm{q}_T}}
& =
\hat{\sigma}_0
H \pqty{Q, \mu}
\sum_q
e_q^2
\mathcal{J}_q \pqty{p_TR, \mu}
\nonumber \\
& \quad \times
\int \frac{b \dd{b}}{2\pi}
J_0 \pqty{q_T b}
x f_q \pqty{x, b, \mu, \zeta}
\overline{S}_{\mathrm{global}} \pqty{b, \mu}
\overline{S}_{\mathrm{cs}} \pqty{b, R, \mu}
, \label{eq:ejet}
\end{align}
where the $\mathcal{J}_q \pqty{p_TR, \mu}$ is the quark jet function and it can be expressed at NLO as \cite{Ellis:2010rwa, Liu:2012sz}:
\begin{align}
\mathcal{J}_q \pqty{p_TR, \mu}
=
1
+
\frac{\alpha_s}{2\pi} C_F
\bqty{\frac{1}{2}\ln[2](\frac{\mu^2}{p_T^2R^2}) + \frac{3}{2} \ln(\frac{\mu^2}{p_T^2R^2}) + \frac{13}{2} - \frac{3\pi^2}{4}}
,
\end{align}
while the remaining ingredients in this factorization formula have been discussed in \cref{sec:collinear}.
The anomalous dimensions of the hard function and the quark TMD PDFs have also been given in \cref{eq:ad-hadrd} and \cref{eq:ad-qTMDPDF}, respectively.
Below we list the anomalous dimensions associated with the soft functions $\overline{S}_{\mathrm{global}} \pqty{b, \mu}$, $\overline{S}_{\mathrm{cs}} \pqty{b, \mu, \nu}$ and the quark jet function $\mathcal{J}_q \pqty{p_TR, \mu}$:
\begin{align}
\gamma_{\mu}^{\overline{S}_{\mathrm{global}}}
& =
\frac{\alpha_s C_F}{\pi}
\pqty{\ln(\frac{\mu^2}{\mu_b^2}) + 2\eta_J}
, \\
\gamma_{\mu}^{\overline{S}_{\mathrm{cs}}}
& =
-\frac{\alpha_s C_F}{\pi}
\ln(\frac{\mu^2}{\mu_b^2R^2})
, \\
\gamma_{\mu}^{\mathcal{J}}
& =
\frac{\alpha_s C_F}{\pi}
\pqty{\ln(\frac{\mu^2}{p_T^2R^2}) + \frac{3}{2}}
.
\end{align}
RG consistency of the differential cross section in \cref{eq:ejet} requires:
\begin{align}
\gamma_{\mu}^H
+
\gamma_{\mu}^q
+
\gamma_{\mu}^{\overline{S}_{\mathrm{global}}}
+
\gamma_{\mu}^{\overline{S}_{\mathrm{cs}}}
+
\gamma_{\mu}^{\mathcal{J}}
=
0
,
\end{align}
which yields:
\begin{align}
\zeta 
= \frac{Q^4 e^{2\eta_J}}{p_T^2}
.
\label{eq:zetaRGCforColl}
\end{align}

Now with $\zeta$ fixed, we consider the in-jet OPEC in lepton-jet production, which involves measuring a hadron inside the jet.
The corresponding factorization formula is given in \cref{eq:injetEC-RL}.
The $\mu$-evolution kernel of the EEC jet function has been given in \cref{e.mu-evolution-J_q} in the back-to-back limit.
Following \cref{eq:TMDFFandTMDFJF}, the anomalous dimension for the EEC jet function in the collinear limit is equivalent to that of the quark TMD FJFs $\mathcal{D}_{h/q} \pqty{z_h, b', \mu, \zeta_J}$:
\begin{align}
\gamma_{\mu}^{\mathcal{D}}
=
\frac{\alpha_s C_F}{\pi}
\pqty{\ln(\frac{\mu^2}{\zeta_J}) + \frac{3}{2}}
.
\end{align}
RG consistency of the differential cross section for OPEC at collinear limits requires:
\begin{align}
\gamma_{\mu}^H
+
\gamma_{\mu}^q
+
\gamma_{\mu}^{\overline{S}_{\mathrm{global}}}
+
\gamma_{\mu}^{\overline{S}_{\mathrm{cs}}}
+
\gamma_{\mu}^{\mathcal{D}}
=
0
,
\end{align}
which yields:
\begin{align}
\zeta \zeta_J
=
Q^4 R^2 e^{2\eta_J}.
\label{e.zeta_zeta_J}
\end{align}

In the center of mass frame for the electron-proton scattering, the momenta of the proton and the produced jet in the light-cone coordinate are:
\begin{align}
P
& =
\bqty{\sqrt{\frac{s}{2}}, 0, 0}
, \\ 
p_J
& =
\bqty{\frac{p_T}{\sqrt{2}} e^{\eta_J}, \frac{p_T}{\sqrt{2}} e^{-\eta_J}, \bm{p}_T}
,
\end{align}
respectively.
Here, square brackets denote vectors written in light-cone coordinates, defined as $v^{\mu} = \bqty{v^+, v^-, \bm{v}_T}$ with $v^{\pm} \equiv \pqty{v^0 \pm v^3}/\sqrt{2}$.
At the Born level, the underlying process is $\gamma^* \pqty{q} + q \pqty{xP} \to q \pqty{p_J}$.
From the momentum conservation, we have $Q^2 \equiv -q^2 = -\pqty{p_J-xP}^2$, and this yields:
\begin{align}
Q^2
=
x \sqrt{s} p_T e^{-\eta_J}
.
\end{align}
We have determined the rapidity scale $\zeta$ appearing in the TMD PDFs via \cref{eq:zetaRGCforColl}, which now can be expressed as $\zeta = x^2 s$. This immediately implies 
$\zeta_J = p_T^2 R^2$ from Eq.~\eqref{e.zeta_zeta_J}.

\newpage
\bibliographystyle{JHEP}
\bibliography{refs}

\end{document}